\documentclass[fleqn,usenatbib]{mnras}

\usepackage{newtxtext,newtxmath}

\usepackage[T1]{fontenc}
\usepackage{booktabs}

\DeclareRobustCommand{\VAN}[3]{#2}
\let\VANthebibliography\thebibliography
\def\thebibliography{\DeclareRobustCommand{\VAN}[3]{##3}\VANthebibliography}

\usepackage{graphicx}	
\usepackage{amsmath}	

\title[GRMHD simulations of black hole accretion variabilities]{GRMHD simulations of black hole accretion variabilities: Implications to hard state X-ray binary transients}

\author[Raha, Mukhopadhyay and Chatterjee]{
Rohan Raha,$^{1}$\thanks{raharohan@iisc.ac.in}
Banibrata Mukhopadhyay,$^{1}$\thanks{bm@iisc.ac.in}
Koushik Chatterjee$^{2}$\thanks{kchatt@umd.edu}
\\
$^{1}$Department of Physics, Indian Institute of Science, 
Bangalore 560012, India\\
$^{2}$Institute for Research in Electronics and Applied Physics, University of Maryland, College Park, MD 20742, USA}

\date{Accepted XXX. Received YYY; in original form ZZZ}

\pubyear{\the\year{}}

\begin{document}
\label{firstpage}
\pagerange{\pageref{firstpage}--\pageref{lastpage}}
\maketitle

\begin{abstract}
Using high-resolution general relativistic magnetohydrodynamic (GRMHD) simulations, we investigate accretion flows around spinning black holes and identify three distinct accretion states. Our results suggest the origin of the complex phenomenology observed across the black hole mass spectrum as the interplay between magnetic and gravitational fields. The magnetically arrested disk (MAD) state, characterized by strong magnetic fields (plasma-$\beta << 1$), exhibits powerful jets, highly variable accretion, and significant sub-Keplerian motion. On the other hand, weakly magnetized disks (plasma-$\beta >> 1$), known as the standard and normal evolution (SANE) state, show steady accretion with primarily winds. An intermediate state bridges the gap between MAD and SANE regimes, with moderate magnetic support (plasma-$\beta \sim 1$) producing mixed outflow morphologies and complex variability.  This unified framework has many implications including its possible connection to extreme variability of GRS~1915+105, particularly in its hard spectral states. It also suggests the possible origin of steady jets of Cyg X-1 and the unusually high luminosities (even super-Eddington based on stellar mass black hole) of HLX-1 without requiring super-Eddington mass accretion rates. Our simulations reveal a hierarchy of timescales that explain the rich variety of variability patterns, with magnetic processes driving transitions between states. Comparing two with three dimensional simulations demonstrates that while quantitative details differ, the qualitative features distinguishing different accretion states remain robust. 
\end{abstract}




\section{Introduction} \label{sec:intro}

The study of accretion flows around black holes has been revolutionized by the ability to perform increasingly sophisticated general relativistic magnetohydrodynamic (GRMHD) simulations in parallel with high-resolution X-ray observations. These numerical experiments have proven crucial for understanding the complex interplay between magnetic fields, gravity, and plasma dynamics that drive accretion and outflow processes.

In the theoretical front, the understanding of black hole accretion has evolved significantly since the $\alpha$-disk model \citep{ShakuraSunyaev1973}, which parameterized angular momentum transport through an effective turbulent viscosity. The discovery of magnetorotational instability (MRI) as the physical mechanism for angular momentum transport \citep{BalbusHawley1991} marked a crucial advance but left open questions about the role of large-scale magnetic fields in driving accretion outflows and jets. This is where GRMHD simulations provided vital insights over the past decades or so \citep[e.g.,][]{Gammie2003, 2006MNRAS.368.1561M, Narayan2012, Porth2019}. We plan to take stock of GRMHD simulations to infer various observational properties of X-ray binaries (XRBs). 

\begin{figure*}
\centering

\includegraphics[width=0.8\textwidth]{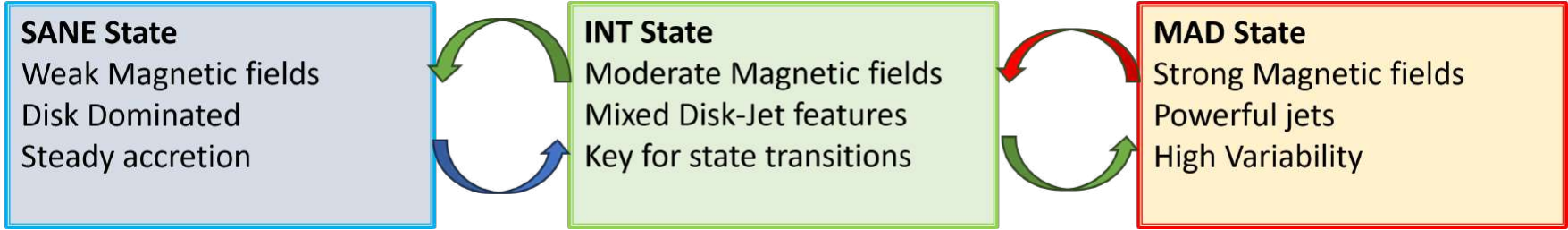}

\caption{ Schematic representation of the three accretion states and their transitions identified in our GRMHD simulations. The SANE state is characterized by weak magnetic fields (plasma-$\beta >> 1$) with low variability in accretion rate and weak jets. The MAD  state features strong magnetic fields (plasma-$\beta << 1$) leading to magnetic stress dominated accretion with strong and powerful jets. The newly identified INT state (plasma-$\beta \sim 1-10$) bridges these extremes with moderate magnetic field strengths and mixed disk-jet characteristics, facilitating transitions between states (indicated by colored arrows). The bi-directional arrows represent the reversible nature of these transitions, with different colors indicating distinct transition pathways. This framework enlightens the possible origin of the complex variability patterns observed in systems like GRS~1915+105, the steady jets in Cyg X-1, and the high luminosities of ULXs like HLX-1.}
\label{fig:state_transition}

\end{figure*}

Two distinct regimes of accretion have emerged from GRMHD studies: the standard and normal evolution (SANE) state where turbulence driven by MRI dominates transport \citep{Narayan2012, Porth2019} with steady accretion, and the magnetically arrested disk (MAD) state where strong poloidal fields significantly influence the accretion flow \citep{Igumenshchev:2003, Narayan2003, Tchekhovskoy2011, raha2024}. The present work focuses also on a crucial third regime - the intermediate state (INT) (sometimes referred to as the ``INSANE'' accretion mode) - which we suggest explains many of the complex transitions observed in X-ray binaries (XRBs). The INT state is practically a new state which we have identified from our simulations that is not described very well in existing literature.

The accretion efficiency in XRBs is fundamentally related to the Eddington luminosity:
\begin{equation}
\label{eqn:Ledd}
L_{\rm Edd} = \frac{4\pi GM m_p c}{\sigma_T}\propto\dot{M}_{Edd}c^2,
\end{equation}
where $G$ is the Newton's gravitational constant, $M$ is the mass of the black hole, $c$ is the speed of light, $m_p$ is the proton mass, $\sigma_T$ is the Thomson scattering cross section and $\dot{M}_{Edd}$ is the Eddington mass accretion rate. To achieve luminosities approaching $L_{Edd}$, as seen in many XRBs, either the flow has high efficiency to convert influx energy to radiation, or it requires a high mass accretion rate. However, MAD/INT/SANE have a strict upper limit of accretion rate, hence influx energy, being non-Keplerian flow. Nevertheless, strong large-scale magnetic fields enable efficient energy extraction through jets and outflows \citep{BisnovatyiKogan1974, BlandfordPayne1982, BlandfordZnajek1977}, apart from angular momentum transport in the disk. While not directly producing radiative efficiency, these magnetic processes can power jets that contribute significantly to the total energy output \citep{Tchekhovskoy2011, McKinney_2012, raha2023}. For rapidly spinning black holes, these magnetic fields can extract rotational energy through the Blandford-Znajek mechanism \citep{BlandfordZnajek1977}. Operating via the Penrose process \citep{Penrose1971}, this mechanism taps into the black hole's spin energy through magnetic field lines twisted by frame-dragging effects near the ergosphere. The ordered magnetic field threading the disk also powers jets through the release of gravitational energy \citep{BlandfordPayne1982}. When the accretion flow acquires strong magnetic fields at large radii, it can develop significant magnetic shear, effectively transporting angular momentum outward \citep{Mondal2018}.

Black hole accretion systems exhibit remarkable diversity in their spectral states and temporal behaviors. On of the black hole XRBs, GRS~1915+105, is the most extraordinary example of this variability. This source displays twelve distinct temporal classes ($\alpha$, $\beta$, $\chi$, etc.), each characterized by unique combinations of spectral properties, timing signatures, and outflow behaviors that challenge our theoretical understanding of accretion physics. The system's complex state transitions occur on timescales ranging from seconds to months, accompanied by sophisticated outflow structures including disk winds (possibly by turbulence or Blandford-Payne mechanism \citep{BlandfordPayne1982}) reaching $1000$ km/s and relativistic jets with Lorentz factors exceeding 2. Particularly intriguing are classes like $\chi$, which exhibit steady hard-state ($\chi$ in fact is canonical low/hard state) emission with persistent radio jets, contrasting sharply with the rapid spectral oscillations of the $\beta$ class and the dramatic ``heartbeat" cycles of the $\rho$ class. The correlation between state transitions and changes in jet activity, combined with theoretical predictions of magnetic flux accumulation driving accretion variability, suggests an underlying connection between magnetic field configurations and the observed accretion states, with different classes potentially representing distinct magnetohydrodynamic equilibria.

The universality of these accretion processes extends across the black hole mass spectrum. While Cyg X-1 maintains remarkably stable hard-state jets with steady radio emission (of $15$ mJy) and consistent jet power (of $10^{37}$ erg/s), its behavior contrasts with both the extreme variability of GRS~1915+105 and the scaled-up temporal patterns observed in HLX-1. The intermediate-mass black hole candidate HLX-1 achieves high luminosities $\sim 10^{40}-10^{41}$ erg/s while maintaining hard spectral states at sub-Eddington accretion rates, presenting a significant challenge to conventional accretion models that typically require substantial beaming or super-Eddington accretion to explain such extreme luminosities. Understanding how magnetic field configurations possibly drive these diverse observational phenomena across different mass scales requires detailed magnetohydrodynamic simulations that can connect the underlying accretion physics to the rich variety of observed behaviors.

In this work, we present systematic GRMHD simulations, in sub-Eddington, sub-Keplerian accretion regime, designed to explore how different initial magnetic field configurations and plasma-$\beta$ parameters (defined as the ratio of gas pressure to the magnetic pressure) influence accretion dynamics. Using the H-AMR \citep{Liska_2022} and HARMPI \citep{Tchekhovskoy:2019} codes, we focus particularly on understanding transitions between different sub-Keplerian accretion states and their possible observational signatures. Our simulations of MAD, INT, and SANE states suggest the possible origin, at least a part of, the observed variety of spectral states, particularly hard spectra, and their transitions. The simulations spanning a wide range of initial conditions allow us to map the relationship between magnetic field geometry and accretion state, which are related to observable properties.

One of the key aspects of our work is the identification and characterization of INT accretion state, which bridges the gap between the well-studied MAD and SANE regimes. 
 
Figure \ref{fig:state_transition} shows a schematic representation of the three distinct accretion states and their transitions. Our framework help understand the complex variability observed in accreting black hole systems, particularly in their sub-Keplerian, advective regime, suggesting their link with magnetic activities.

The paper is organized as follows. Section~\ref{sec:results} presents our 3D simulation results, focusing on the temporal evolution of accretion regimes (motivated by the 2D parameter survey detailed in Appendix~\ref{sec:simulations}). Section \ref{sec:comparison} compares the results of 3D simulations reported in Section \ref{sec:results} with those of high resolution 2D simulations. Section~\ref{sec:conclusions} summarizes our conclusions and discusses the implications of our results for understanding accretion physics across the black hole mass spectrum.

\section{3D High Resolution Models} \label{sec:results}

\begin{figure}
\centering

\begin{minipage}[c]{0.5\textwidth}
  \includegraphics[width=1.2\textwidth,height=6cm,keepaspectratio]{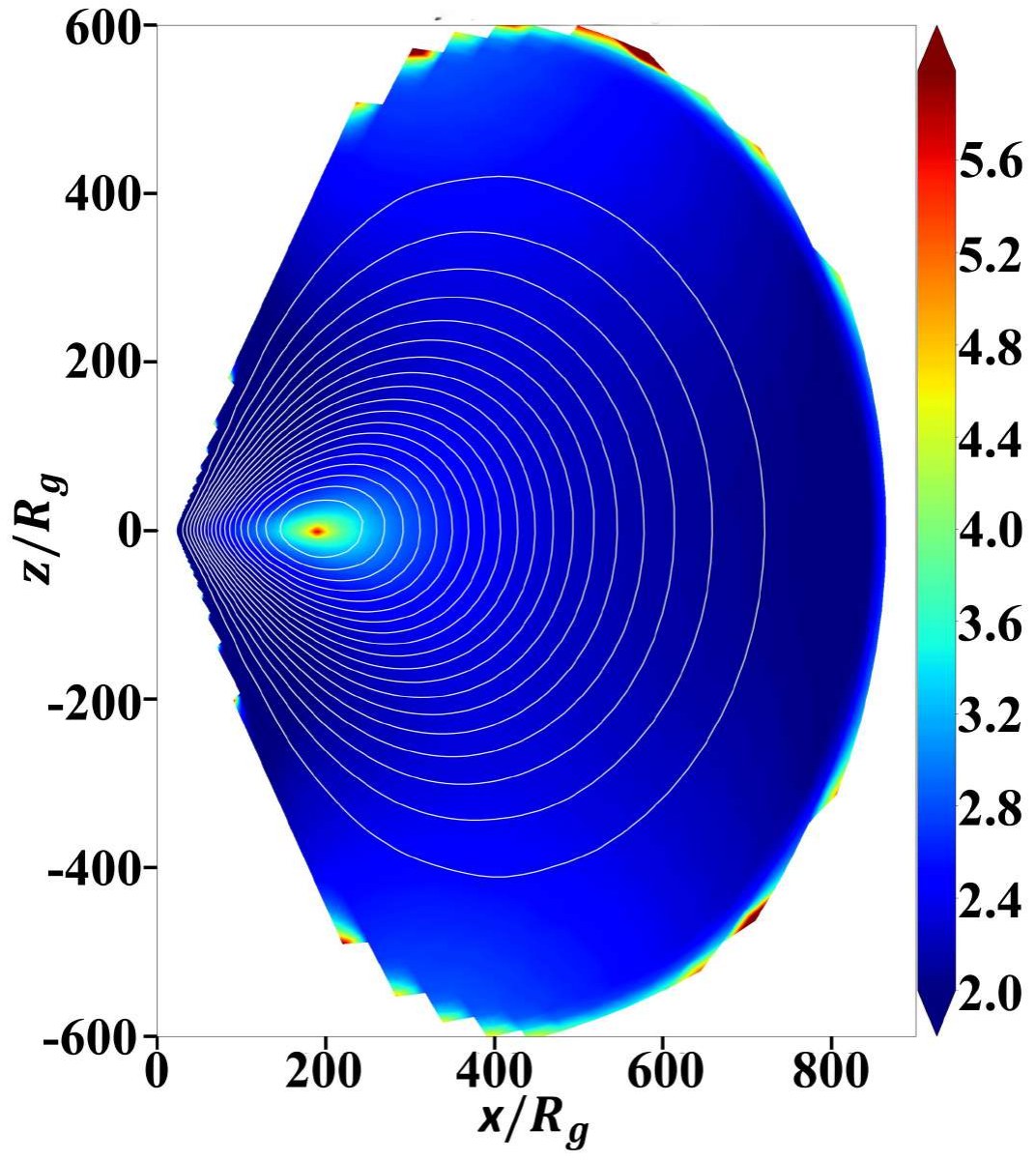}

   \hspace{4cm}(a) MAD
\end{minipage}
\vspace{0.3cm}
\begin{minipage}[c]{0.5\textwidth}
  \includegraphics[width=1.2\textwidth,height=6cm,keepaspectratio]{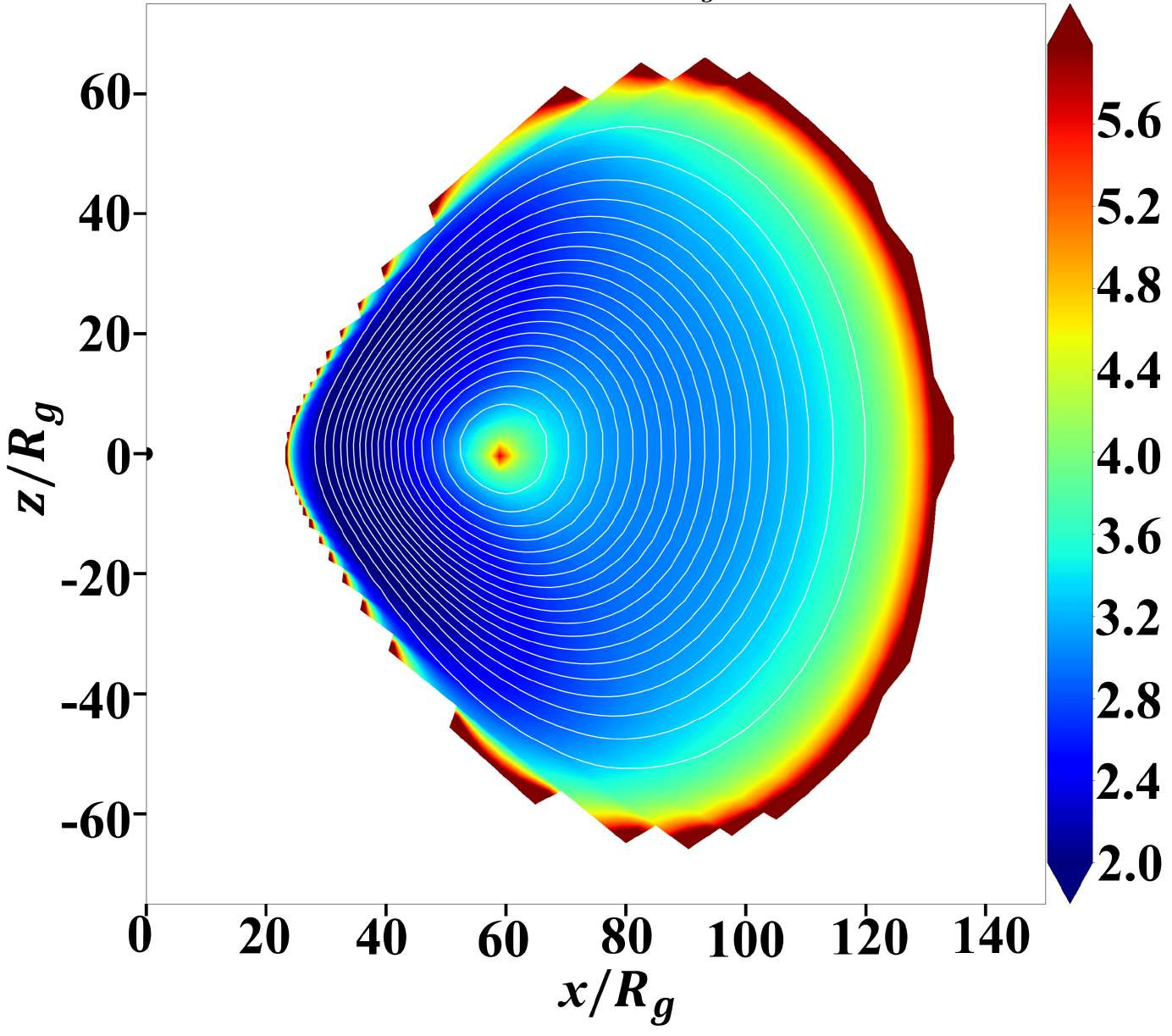}

   \hspace{4cm}(b) INT
\end{minipage}
\vspace{0.3cm}
\begin{minipage}[c]{0.5\textwidth}
  \includegraphics[width=1.2\textwidth,height=6cm,keepaspectratio]{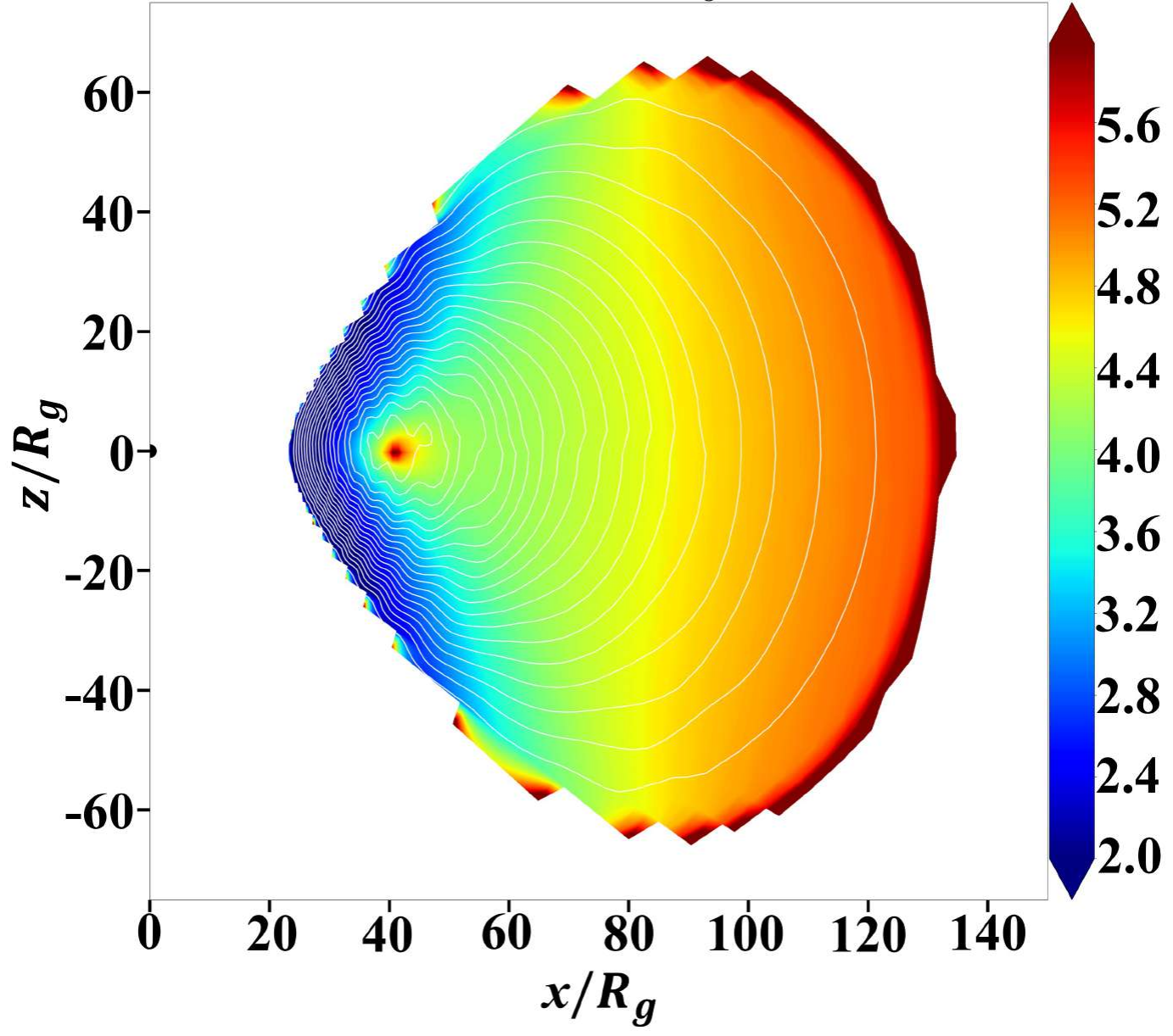}
  \centering
   \hspace{4cm}(c) SANE
\end{minipage}

\caption{Initial magnetic field configurations for our three primary simulations. The color contours represent the logarithmic plasma-$\beta$, while the white lines show the magnetic field lines: (a) MAD configuration features a strong magnetic field structure near the equatorial plane with very low plasma-$\beta$ distribution and a very large initial torus. (b) INT configuration creates a smaller disk with relatively higher plasma-$\beta$. (c) SANE configuration shows a simpler poloidal field pattern with even higher plasma-$\beta$ and a very low density of magnetic field lines, and smaller loops compared to MAD and INT configurations.}
\label{fig:beta0}
\end{figure}

In this section, our analysis focuses on high-resolution 3D GRMHD simulations performed using H-AMR. Based on systematic exploration of different initial magnetic field 
configurations and plasma-$\beta$ values (detailed in Appendix \ref{sec:simulations}), 
we identify three configurations that represent distinct accretion 
regimes. The results may have connection to distinct accretion states observed in black hole systems. Out of all simulations described in {Appendix \ref{sec:simulations} (see Table \ref{tab:models}), we examine three representative configurations: P2B100, PLB100, and P1B100, which are respectively MAD, INT and SANE states as will be meant throughout this section. While our models focus on gas accretion from rotating tori, substantial diversity arises in other accretion modes due to differences in disk size, angular momentum, and the initial magnetic field configuration. Accretion from larger tori, as well as low–angular-momentum Bondi-type inflows, often leads to magnetically saturated regions extending to $r \gtrsim 3000\,r_g$. In such cases, ejected magnetic flux tubes can propagate largely unimpeded through the infalling gas, accompanied by strong variations in jet efficiency and morphology \citep{Lalakos2022, ChoNarayan2025, Galishnikova2025}. Alternative magnetic field configurations, such as initially pure toroidal fields \citep{Komissarov:2006}, drive in situ poloidal field generation and produce episodes of incoherent jet and wind activity \citep{Liska2020}. Despite this enhanced variability, particularly in jet morphology and efficiency, their time-averaged global properties within a few tens of $r_g$ are expected to be broadly similar to those of our chosen models.

All physical quantities presented here are computed through density-weighted shell averaging over $\theta$ and $\phi$ directions. These configurations are selected based on their ability to reproduce key observational features: P2B100 demonstrates the characteristics of magnetically arrested accretion with strong jet formation, PLB100 captures the complex variability of intermediate states, and P1B100 represents a weakly magnetized flow presumably dominated by MRI. The equations for calculating density-weighted averages of physical quantities - including velocity fields, magnetic field strength, plasma-$\beta$, scale height, and various efficiency measures - are provided in Appendix \ref{sec:2dresults}.

Later, in Section \ref{sec:comparison}, we compare our 3D analysis with 2D results of P2B100 (MAD state), PLB100 (INT state), and P1B100 (SANE state) configurations. This comparison demonstrates that while 3D effects introduce some differences, some of the qualitative features and conclusions drawn from our 2D analysis also remain robust.

To better understand how different magnetic field configurations evolve into distinct accretion states, we show in Figure \ref{fig:beta0} the initial magnetic field topologies for our three primary configurations. The figure displays the initial density distributions overlaid with magnetic field lines for the MAD, INT, and SANE setups in our 3D simulations.

\subsection{Magnetic States  - Understanding Flow Properties }

\begin{figure}
\centering
    \includegraphics[width=0.5\textwidth]{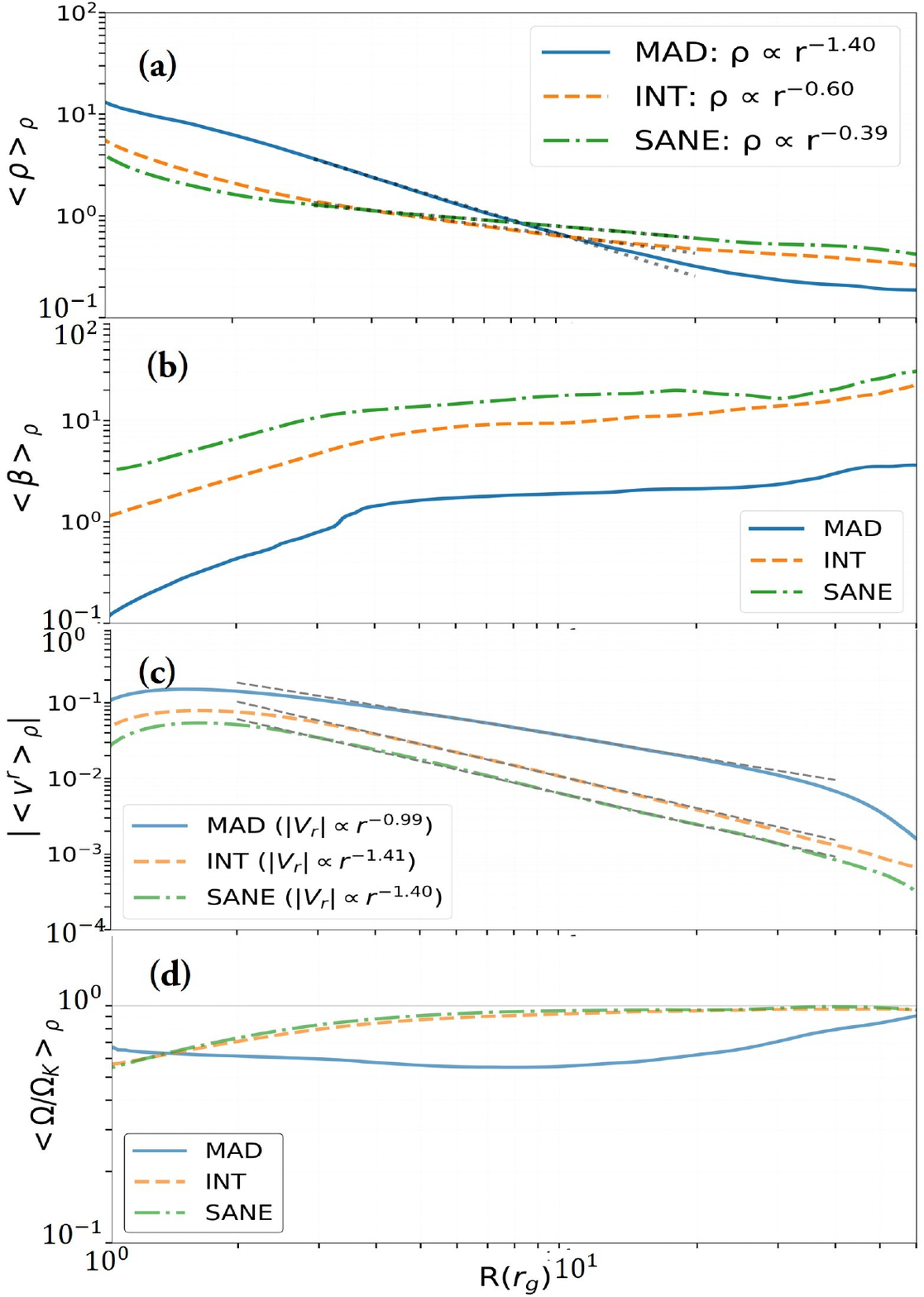}
    \caption{(a) Averaged density as a function of distance from black hole showing power-law relations; power-law indices are computed in the inner region $3-20 r_g$ where the simulation is well within converged and steady zone. (b) Shell averaged  plasma-$\beta$ as a function of distance from black hole for 3D simulations. (c) Shell averaged radial velocity as a function of distance from black hole in units of speed of light for 3D simulations. (d) Ratio of angular velocity to the Keplerian angular velocity as a function of distance from the black hole for 3D simulations.}
    \label{fig:rho_pbeta_vr_om}
\end{figure}

The radial density profiles (Figure \ref{fig:rho_pbeta_vr_om}a) exhibit distinct power-law scalings characteristic of different accretion states. We compute the power-law slopes over the inner region $r = 3 - 20\,r_g$, where the simulations have roughly converged to steady state. The MAD configuration shows the steepest profile, $\rho \propto r^{-1.40}$; the INT state is shallower, with $\rho \propto r^{-0.6}$; and the SANE configuration exhibits the flattest profile, $\rho \propto r^{-0.39}$.

These differences arise from the combined effects of magnetic pressure support, rotation, and radial flow dynamics. In the MAD state, strong magnetic fields (Figure \ref{fig:rho_pbeta_vr_om}b) enforce disk plasma-$\beta \ll 1$ near the horizon, providing substantial magnetic pressure support against gravity \citep{Chatterjee_2022}. As a result, the flow becomes strongly sub-Keplerian, centrifugal support is reduced, and the radial infall speed increases, producing a steep density gradient. In contrast, SANE disks are primarily rotation-supported: nearly Keplerian motion slows the inward radial transport of material, leading to a much shallower density profile. The INT state occupies an intermediate regime, with plasma-$\beta \sim 1$ and magnetic, thermal, and centrifugal forces contributing comparably to the force balance.

These trends are reflected directly in the radial velocity profiles (Figure \ref{fig:rho_pbeta_vr_om}c). MAD maintains significantly higher infall velocities, with $|\langle v^r \rangle_\rho| \sim 0.1c$ at the horizon and a shallow radial scaling $|v^r| \propto r^{-0.99}$, whereas SANE exhibits lower infall speeds, $|\langle v^r \rangle_\rho| \sim 0.03c$, and a steeper decline $|v^r| \propto r^{-1.40}$. The rotation profiles (Figure \ref{fig:rho_pbeta_vr_om}d) show that MAD flows are highly sub-Keplerian, with $\langle \Omega/\Omega_K \rangle_\rho \sim 0.55$ at the horizon, while SANE remains close to Keplerian, $\langle \Omega/\Omega_K \rangle_\rho \sim 0.95$. The higher radial velocities in MAD imply shorter inflow timescales: material spends less time at intermediate radii before plunging into the black hole, suppressing mass accumulation and steepening the density profile. The reduced centrifugal support in MAD, driven by strong magnetic pressure gradients, enables rapid radial infall and reinforces the steep density gradient.

The mass flow rates (Figure \ref{fig:mdot_all}) provide complementary evidence for these dynamical differences. The total mass accretion rate $\dot{M}_{\rm tot}$ remains approximately constant with radius in all three states, confirming that the simulations have reached a quasi-steady state within $30-40\,r_g$. However, the inflow and outflow components differ markedly across accretion regimes.
The inflow mass rate $\dot{M}_{\rm in}$ exhibits two key differences across states: the absolute magnitude, and the radial scaling. 
In our MAD model, the inflow rate $\dot{M}_{\rm in}$ reaches a higher absolute value at the horizon and increases gradually with radius ($\dot{M}_{\rm in} \propto r^{0.39}$), indicating efficient inward transport with minimal mass accumulation at intermediate radii. This combination naturally produces a steep density profile. In contrast, the SANE flow exhibits both lower inflow rates at the horizon and a steeper radial scaling ($\dot{M}_{\rm in} \propto r^{0.80}$), consistent with material accumulating at larger radii as it slowly spirals inward through turbulent transport, yielding a flatter density profile. The INT state again shows intermediate behaviour, both in absolute value and scaling ($\dot{M}_{\rm in} \propto r^{0.67}$). 

Outflow rates further distinguish the accretion states and reveal important differences in outflow morphology. At small radii ($r \lesssim 10\,r_g$), MAD exhibits \emph{lower} absolute outflow rates than SANE despite having stronger magnetic fields. This occurs because MAD's powerful, wide collimated jets occupy a large fraction of the polar solid angle, leaving a reduced disk wind region. Since the shell-integrated $\dot{M}_{\rm out}$ predominantly measures disk wind contributions rather than the highly collimated jet itself, MAD shows lower total outflow rates at these inner radii. In contrast, SANE lacks strong jets, allowing turbulent MRI-driven winds to emerge from a larger azimuthal extent of the disk, producing higher shell-integrated $\dot{M}_{\rm out}$ despite individually weaker flow speeds.

Outflow rates further distinguish the accretion states and reveal important differences in outflow morphology. Outflow rates further distinguish the accretion states. Generally, we see that the stronger the magnetic field, the larger the radius where the outflow rate becomes important (i.e., where $\dot{M}_{\rm out}\approx \dot{M}_{\rm tot}$), as the jet becomes progressively more powerful and grows wider at its base. At small radii ($r \lesssim 10\,r_g$), MAD exhibits \emph{lower} absolute outflow rates than SANE despite having stronger magnetic fields. This reflects MAD's highly collimated wide jet structure at inner radii, suppressing broader wind regions. In contrast, SANE produces more spatially distributed, turbulent mass-loaded outflows/winds even at small radii due to less dominance of jets compared to MAD, resulting in higher shell-integrated $\dot{M}_{\rm out}$ despite weaker outflow velocities. The steep scaling in case of MAD occurs due to the presence of regularly ejected magnetic flux tubes, which power intermittent disk winds that carry out significant amounts of disk material \citep{Chatterjee_2022}.

\begin{figure}
\centering
 \includegraphics[width=0.5\textwidth]{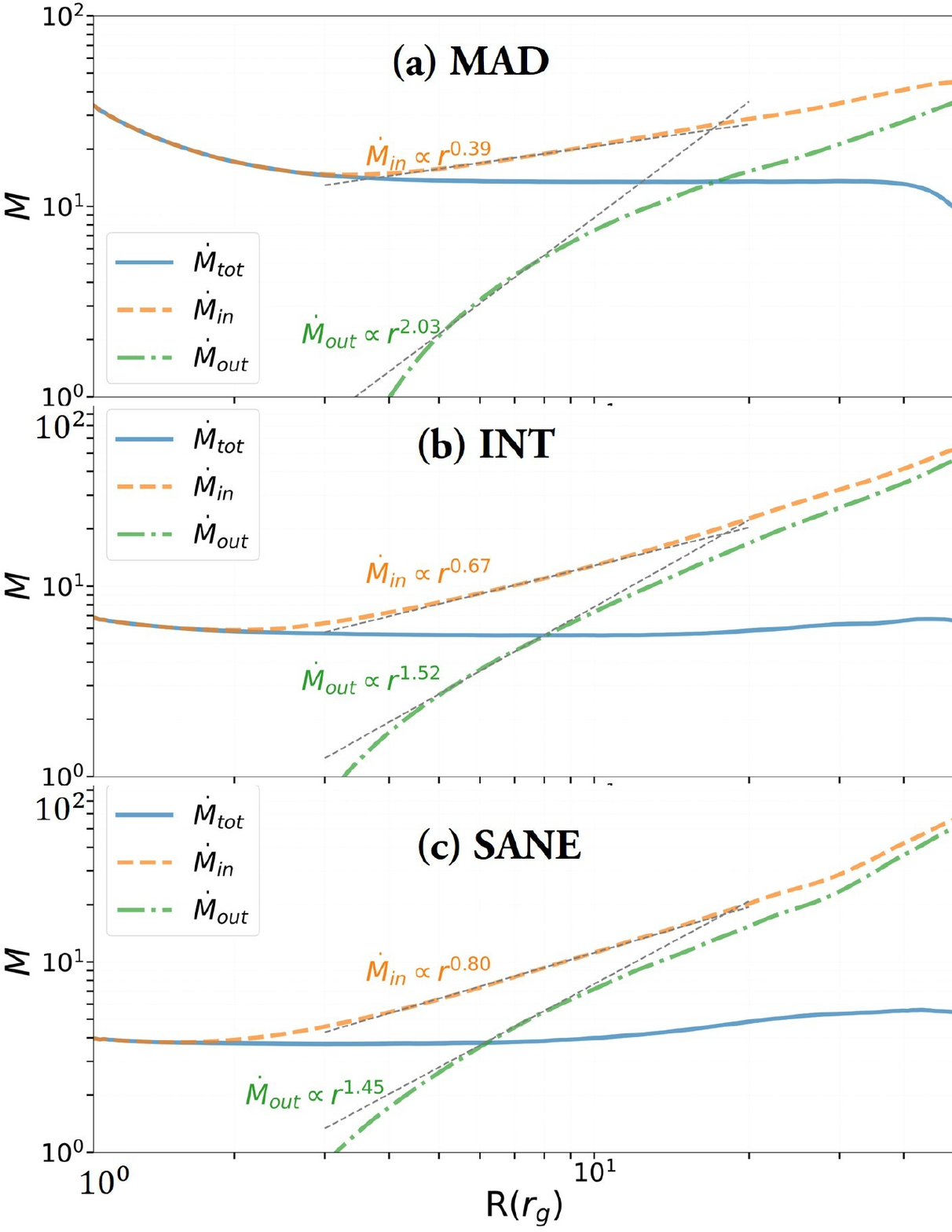}
\caption{Mass flow rates as functions of radius for the three accretion states from 3D simulations. Each panel shows the total mass accretion rate $\dot{M}_{\rm tot}$ (solid blue), inflow rate $\dot{M}_{\rm in}$ (dashed orange), and outflow rate $\dot{M}_{\rm out}$ (dash-dotted green). Gray dashed lines indicate power-law fits in the region $r = 3-20r_g$. (a) MAD state exhibits gradual inflow accumulation ($\dot{M}_{\rm in} \propto r^{0.39}$) and steep outflow scaling ($\dot{M}_{\rm out} \propto r^{2.03}$), characteristic of efficient magnetically-driven jet launching. (b) INT state shows intermediate behavior with $\dot{M}_{\rm in} \propto r^{0.67}$ and $\dot{M}_{\rm out} \propto r^{1.52}$. (c) SANE state displays steeper inflow accumulation ($\dot{M}_{\rm in} \propto r^{0.80}$) and shallower outflow scaling ($\dot{M}_{\rm out} \propto r^{1.45}$), reflecting mass-loaded winds. The relatively constant $\dot{M}_{\rm tot}$ across all states confirms quasi-steady-state conditions. }
\label{fig:mdot_all}
\end{figure}

The density profiles also help understand the observed spectral properties. The steeper power-law slope in MAD means that while their densities might be comparable or slightly higher to SANE very close to the black hole ($r < 5r_g$), they fall off much more rapidly with radius. At the intermediate regions ($r \sim 10-100r_g$), MAD actually has lower densities than SANE due to its steep profile. These lower densities at intermediate radii seem to be responsible for lower optical depths and harder spectra. The key is the integrated column density through the emission region.

\begin{figure*}
\centering
\includegraphics[width=0.33\textwidth]{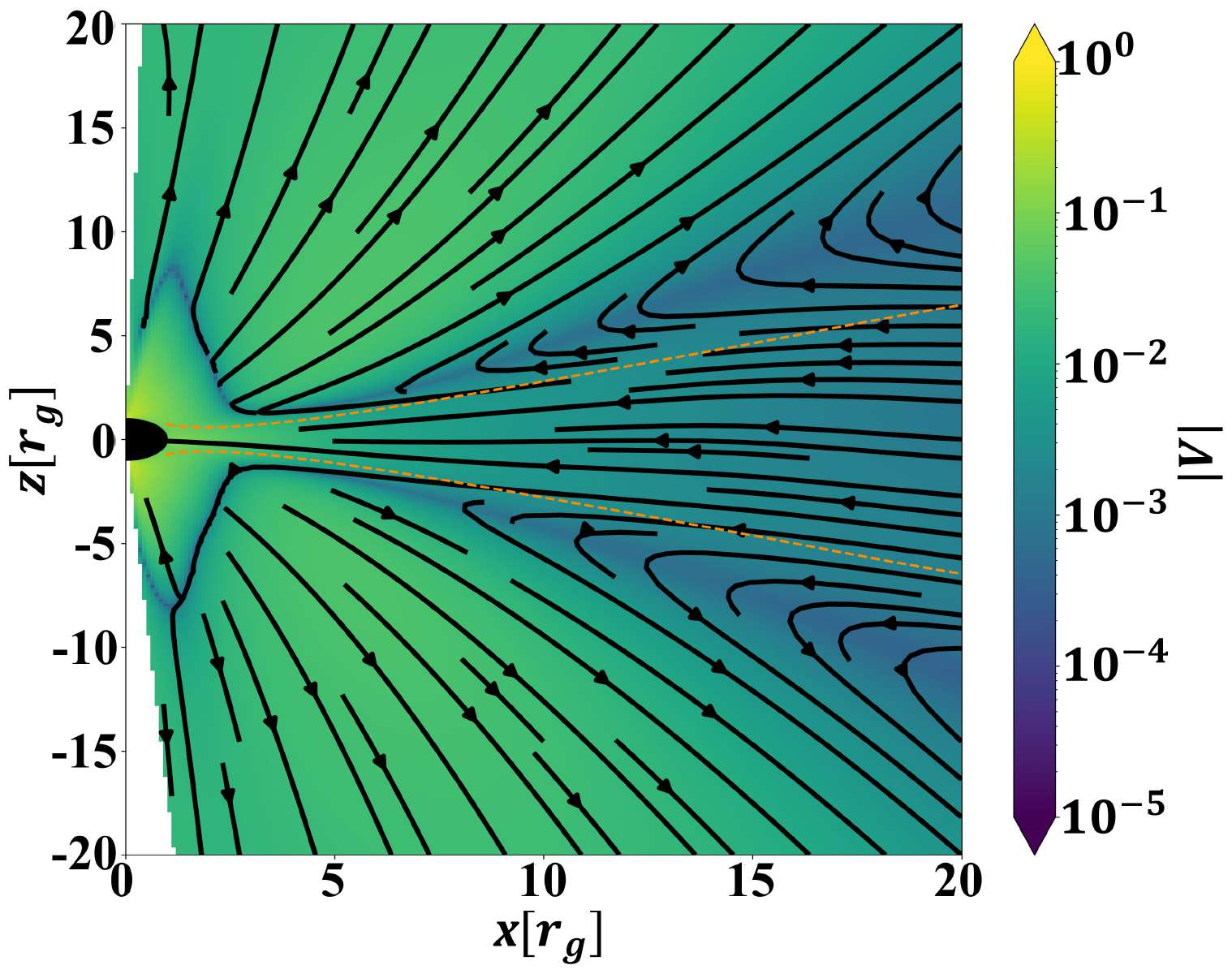}
\includegraphics[width=0.33\textwidth]{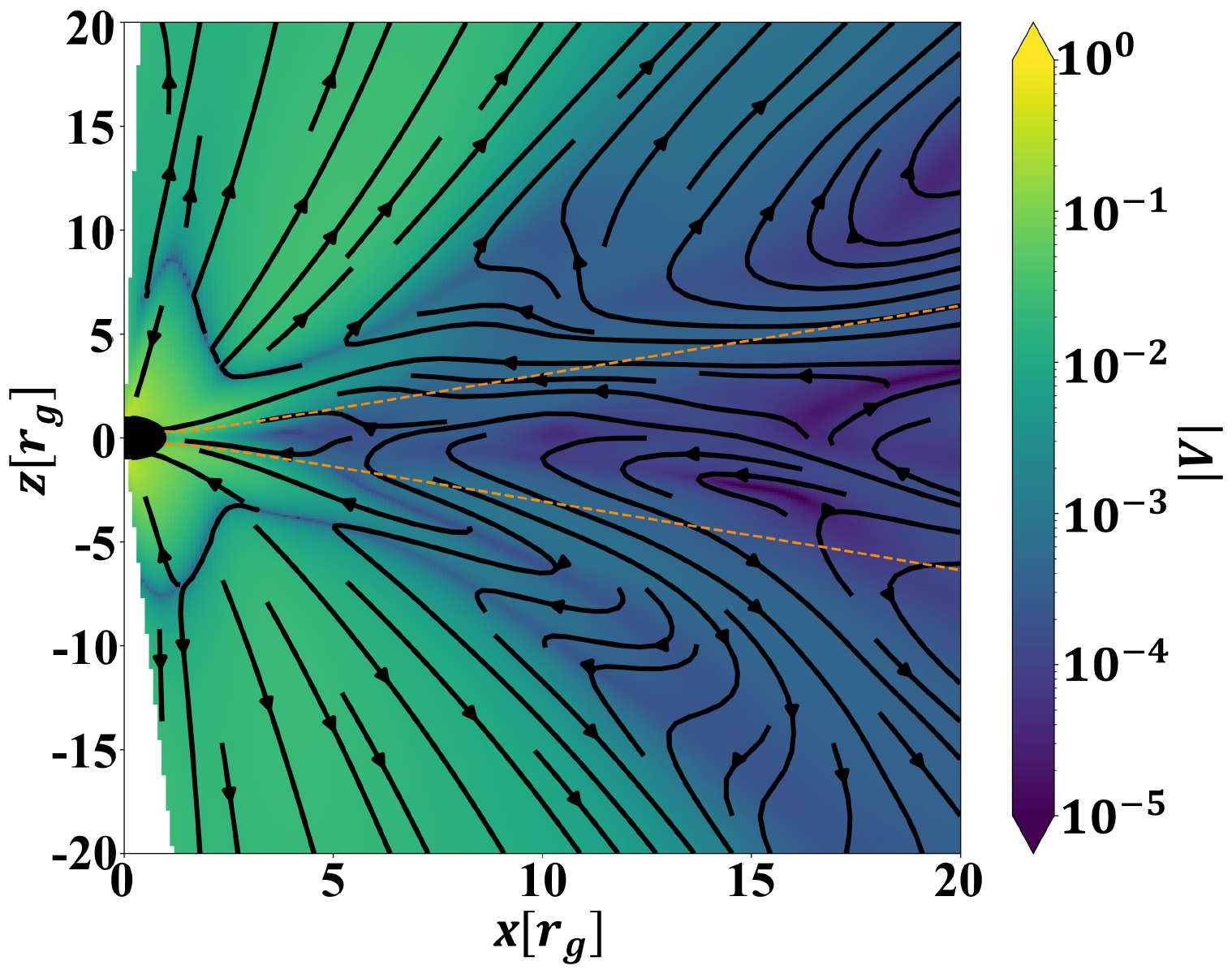}
\includegraphics[width=0.33\textwidth]{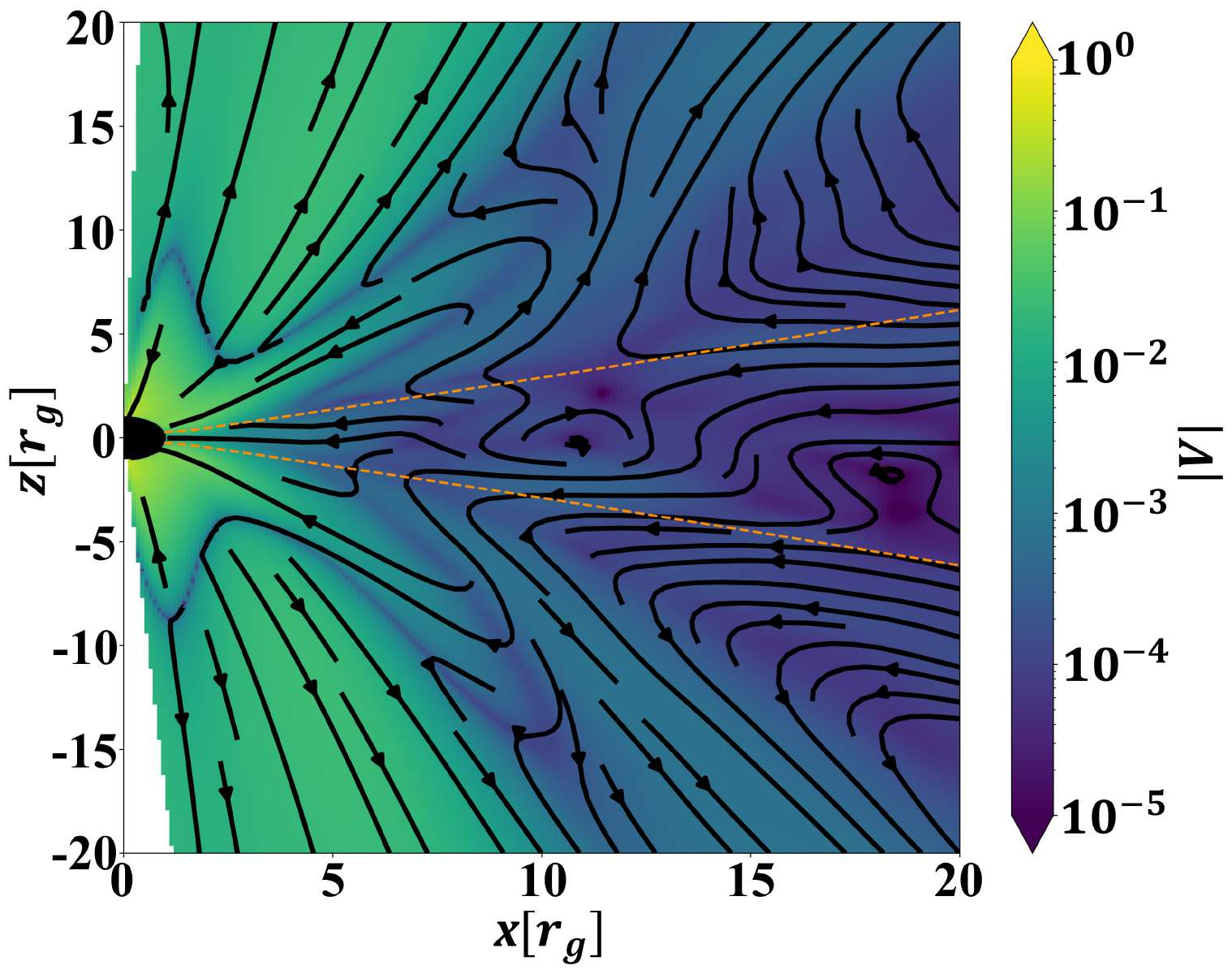}
\caption{Velocity contours for 3D simulations, time averaged from $t=20,000 r_g/c$ to $t=25,000 r_g/c$. The colors represent velocity magnitude and arrows show directions of velocity. The dashed orange line shows the disk boundary calculated from the aspect ratio. Left: MAD. Middle: INT. Right: SANE.}
\label{fig:velocity_fields}
\end{figure*}

The velocity profiles in Figure \ref{fig:velocity_fields} show a clear evolution from highly turbulent flow patterns in SANE to increasingly organized structures in INT and MAD, with SANE displaying complex vortical structures and velocity contours while MAD exhibits smooth, predominantly radial outflow patterns. Inside the disk region, SANE shows strong turbulent mixing with multiple recirculation zones, INT displays intermediate organization, and MAD demonstrates nearly laminar radial motion, while outside the disk region, all three show outward flow but with decreasing complexity from SANE's turbulent streamlines to MAD's uniform, parallel flow characteristic of a steady wind or jet. This progression suggests an evolution from an active, turbulent accretion state (SANE) through an intermediate regime (INT) to an organized outflow-dominated state (MAD). The jet region also increases progressively from SANE to INT to MAD. The jet in turn drags the matter coming out of the disk to form the winds that are seen at the edge of the disk region. The winds are more organized in INT and MAD but are turbulent in SANE as some of the matter from the wind is also seen to change direction as it goes inwards. In the following, we discuss the configurations of MAD (Figures \ref{fig:density_contours_MAD_xz_3D}, \ref{fig:density_contours_MAD_xy_3D}), INT (Figure \ref{fig:density_contours_INT_3D}), and SANE (Figure \ref{fig:density_contours_SANE_3D}) in detail.

\begin{figure*}
\centering
\includegraphics[width=0.45\textwidth]{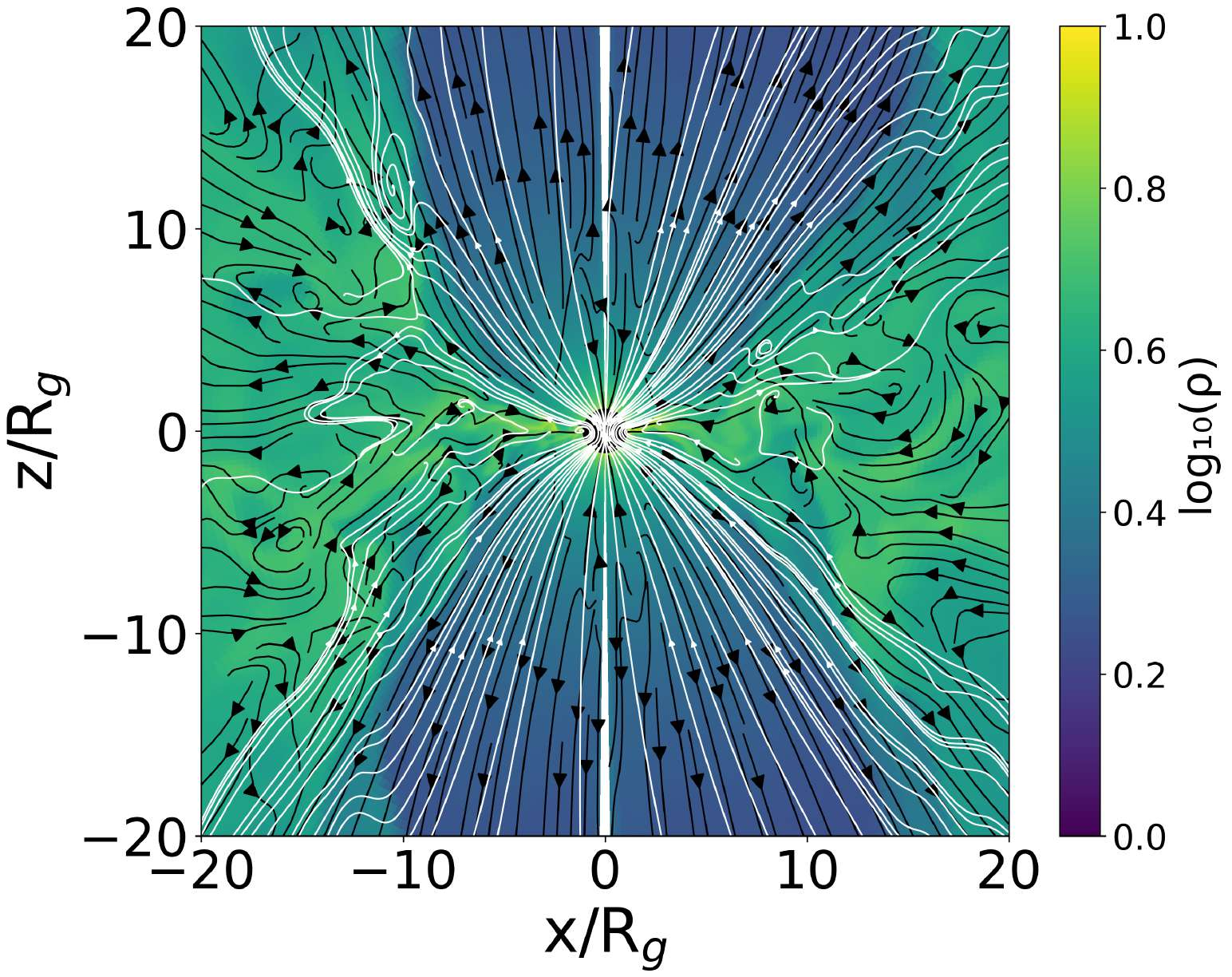}
\includegraphics[width=0.45\textwidth]{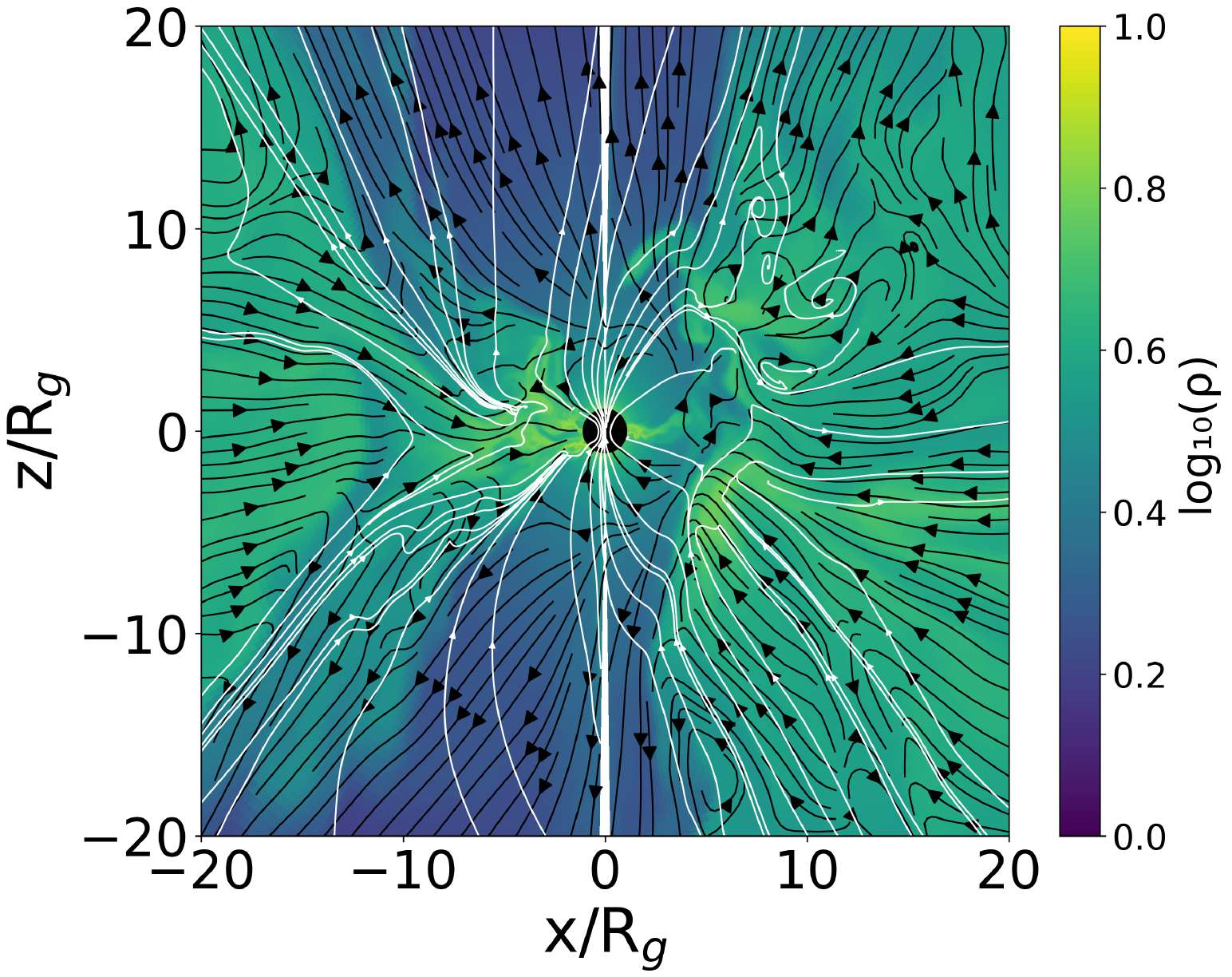}
\caption{Instantaneous density color contours of 3D simulations in $x$-$z$ plane for MAD configuration. The color shows density values in logarithmic scale, the white lines are the magnetic field lines and the black arrows indicate velocity directions: (left) Accretion of matter and magnetic field with wide jet launching regions along the polar axes at $t=23340\,r_g/c$; (right) formation of strong magnetic barriers in the MAD state due to accumulation of magnetic flux which stops matter from accreting further which is characteristic of MAD states at $t=23690\,r_g/c$. An animation of this figure is available online at \url{https://www.youtube.com/watch?v=SCg3zeG5iUI&list=PLHH19kmpHHtG6ONUUhqppc5gKlESPKLG-}. The movies depict time evolution of logarithmic density contours in the side view (Cartesian ($x$, $z$) plane) for MAD simulation of $a = 0.998$ from $t=0$ to $t=25,000\,r_g/c$. The density contour starts with the initial Fishbone--Moncrief torus configuration and matter accretion is initiated. Accretion onto the black hole is observed via a thin funnel around the mid-plane. Due to high magnetic field, repeated accumulation and eruption of matter and magnetic flux is observed.}
\label{fig:density_contours_MAD_xz_3D}
\end{figure*}

\begin{figure*}
\centering
\includegraphics[width=0.45\textwidth]{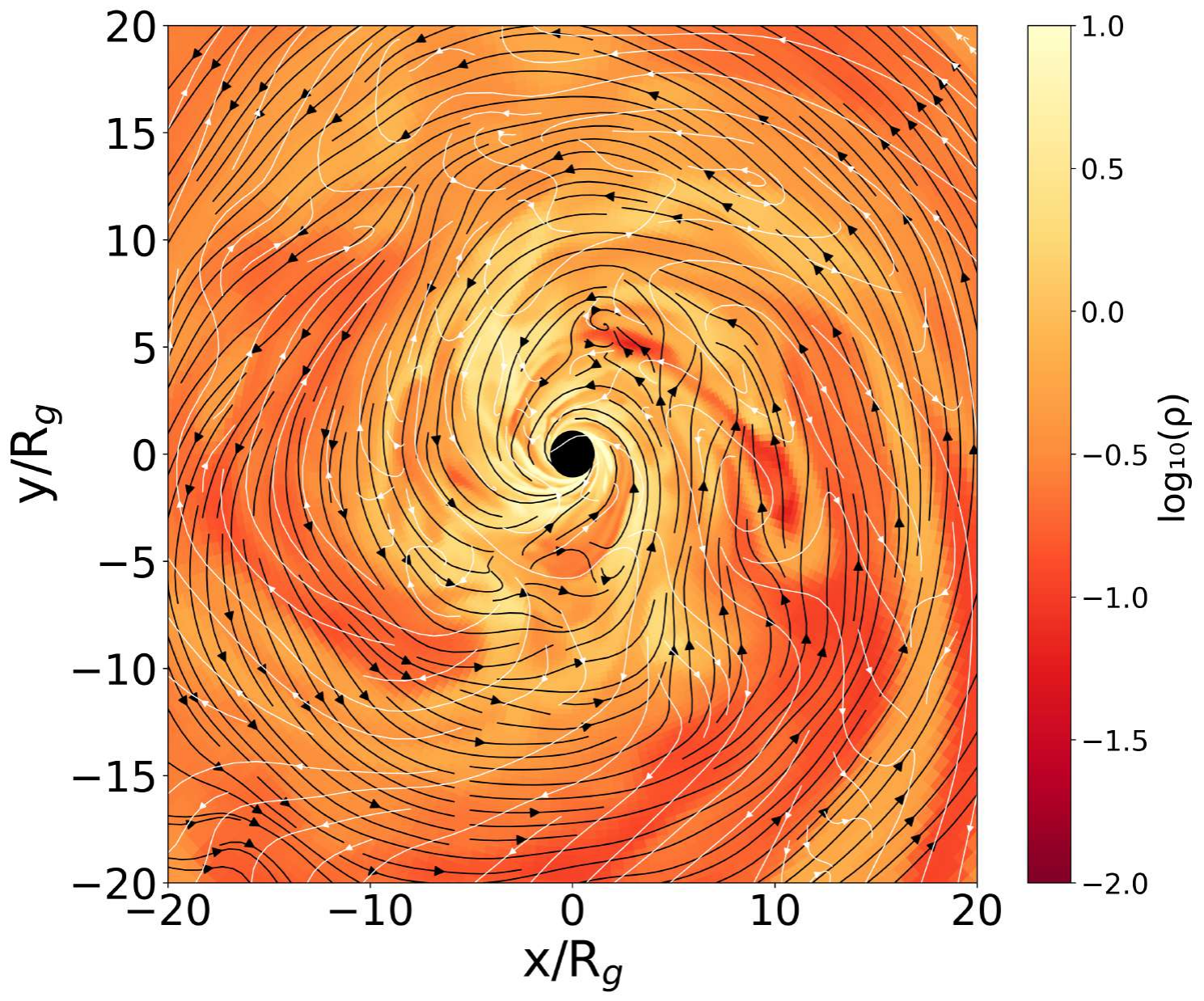}
\includegraphics[width=0.45\textwidth]{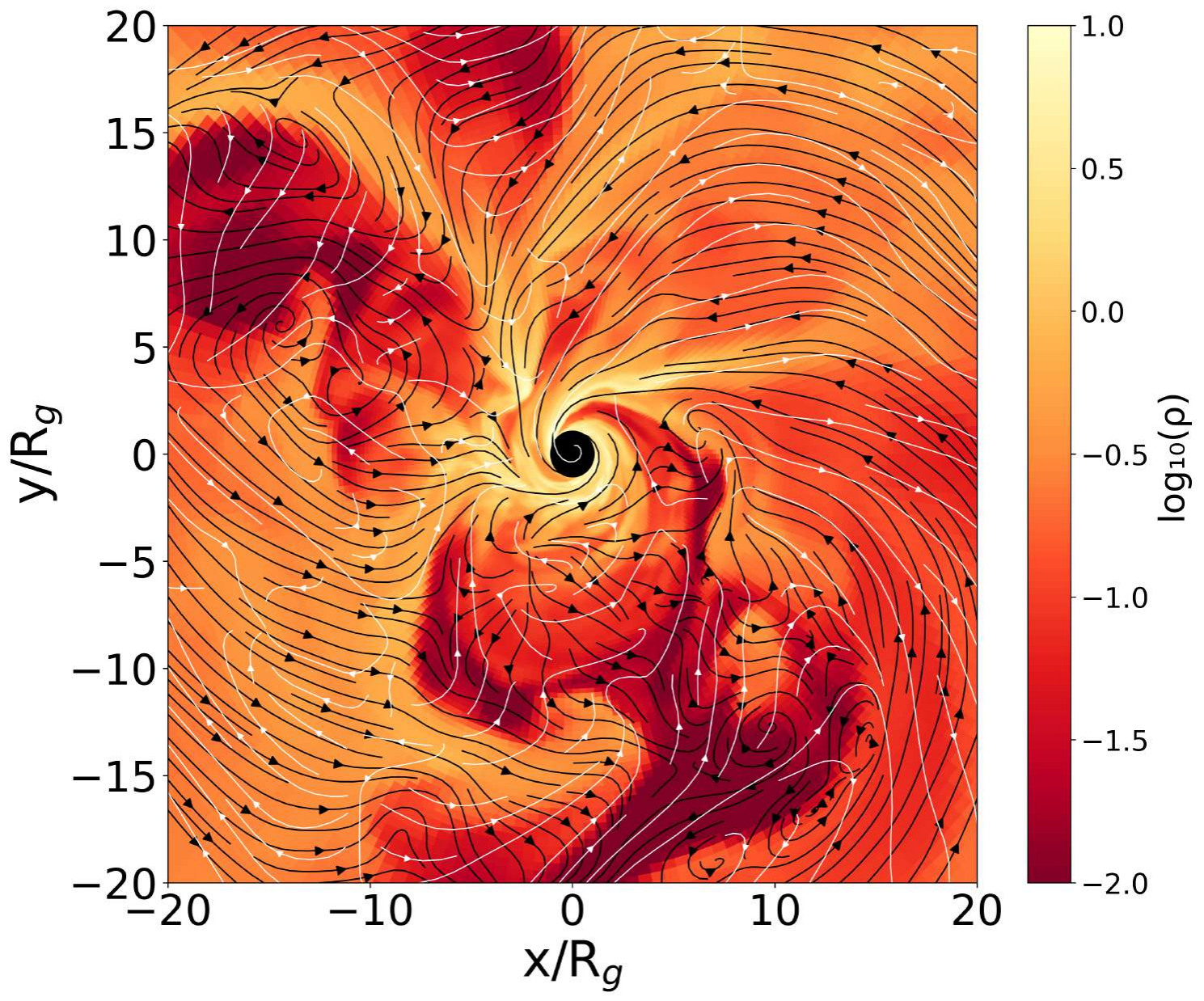}
\caption{Same as Fig.~\ref{fig:density_contours_MAD_xz_3D} except for top (equatorial plane) view in the $x$-$y$ plane: (left) Accretion of matter and magnetic field as seen from top view at $t=23340\,r_g/c$; (right) eruption of matter due to accumulation of magnetic flux as seen from darker regions marked with low matter density at $t=23690\,r_g/c$. An animation of this figure is available online at \url{https://www.youtube.com/watch?v=dQp8hCBks7Y&list=PLHH19kmpHHtG6ONUUhqppc5gKlESPKLG-&index=4}. The movie depicts a top view of accretion and ejection of matter with time.}
\label{fig:density_contours_MAD_xy_3D}
\end{figure*}

\begin{figure*}
\begin{minipage}{0.48\textwidth}
    \includegraphics[width=\textwidth]{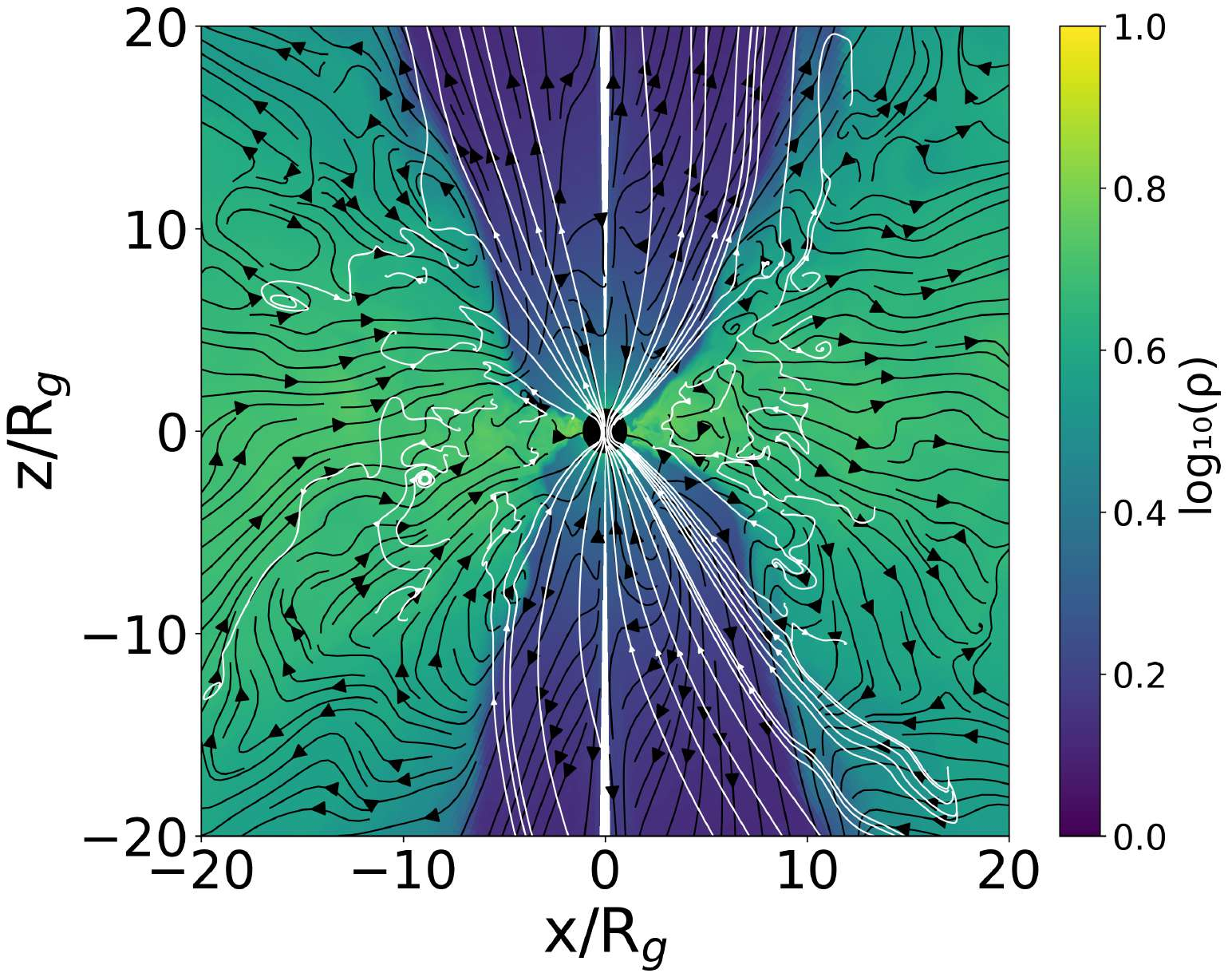}
    \caption{Same as Figure \ref{fig:density_contours_MAD_xz_3D} except for INT state at $t=24020 r_g/c$. Formation of a weaker barrier of magnetic flux in the INT state resulting in steadier accretion of matter compared to MAD state. [See our high resolution movie of density evolution with time on YouTube playlist: \url{https://www.youtube.com/watch?v=NdFmP1kMgBk&list=PLHH19kmpHHtG6ONUUhqppc5gKlESPKLG-&index=3}. Movie depicts steadier accretion with lesser eruptions compared to MAD state.] \\
(An animation of this figure is available.)}
    \label{fig:density_contours_INT_3D}
\end{minipage}
\hfill
\begin{minipage}{0.48\textwidth}
    \includegraphics[width=\textwidth]{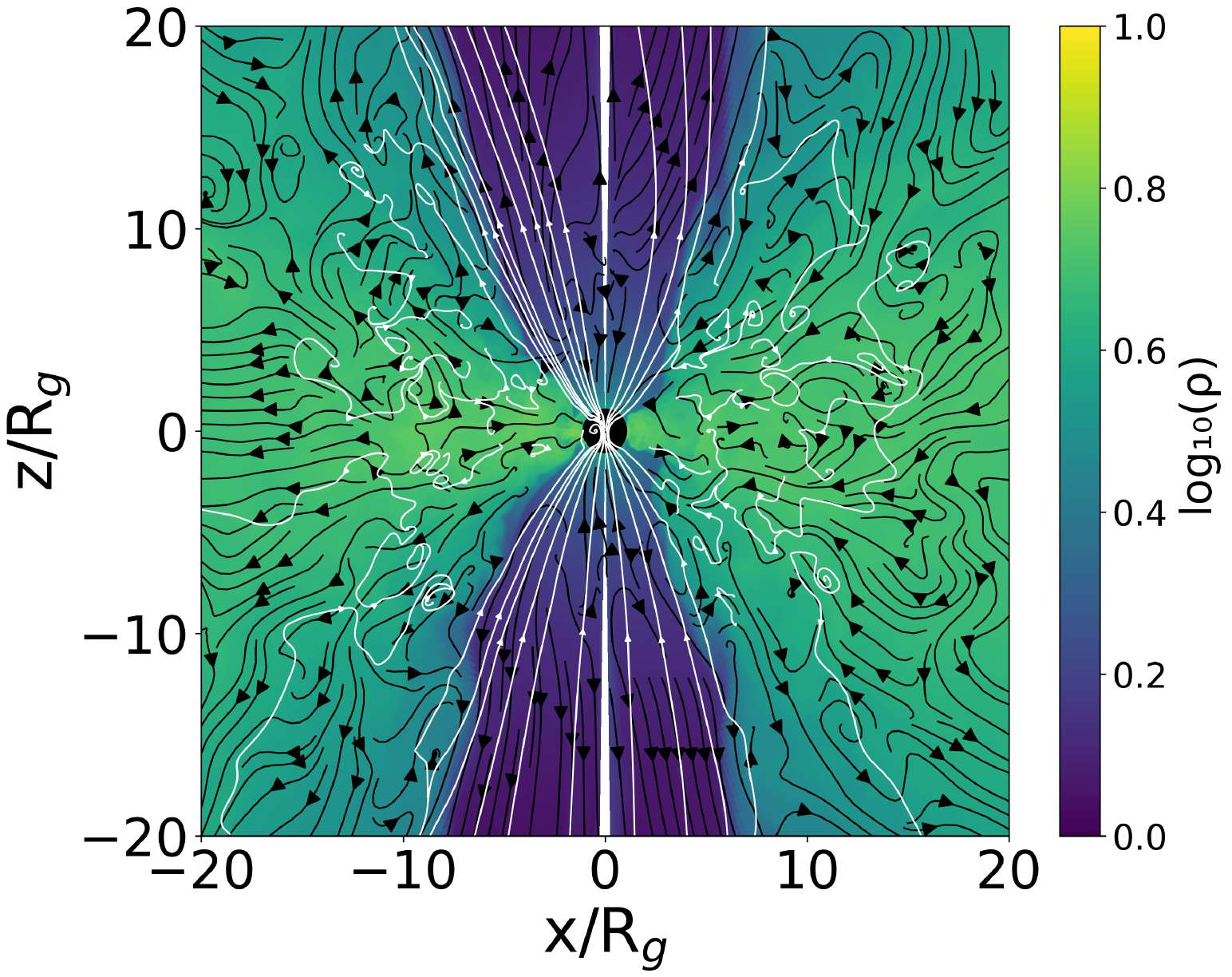}
        \caption{Same as Figure \ref{fig:density_contours_MAD_xz_3D} except for SANE state at $t=22620 r_g/c$. Steady accretion of matter with turbulence due to formation of eddies seen from velocity streamlines in the disk region. [See our high resolution movie of density evolution with time on YouTube playlist: \url{https://www.youtube.com/watch?v=-Q8JFpArDa8&list=PLHH19kmpHHtG6ONUUhqppc5gKlESPKLG-&index=2}. Movie shows steady accretion with time with almost no eruptions unlike MAD state.]\\
(An animation of this figure is available.)}
    \label{fig:density_contours_SANE_3D}
\end{minipage}
\end{figure*}

\begin{figure*}
\begin{minipage}{0.48\textwidth}
    \includegraphics[width=\textwidth]{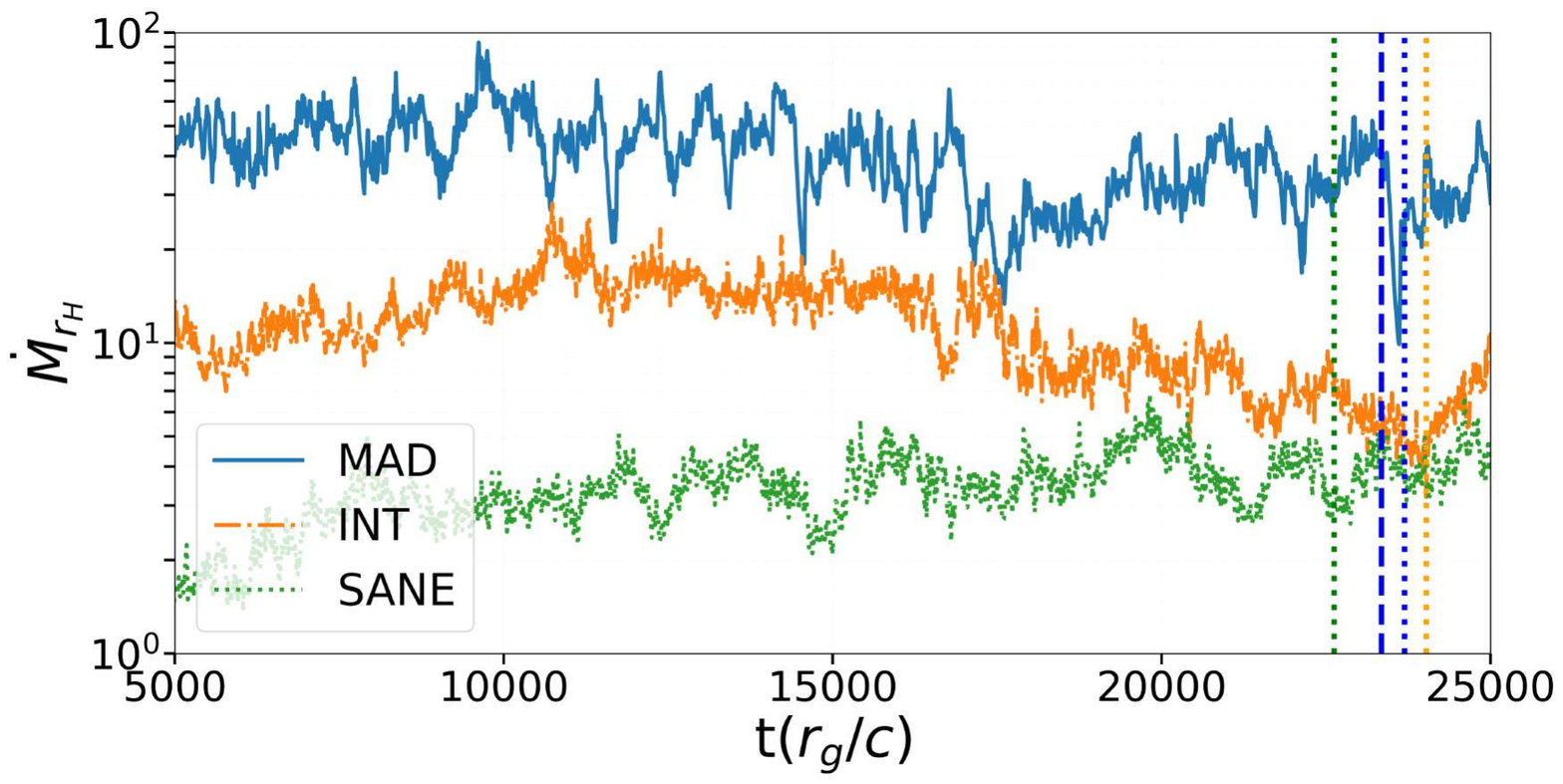}
    \vspace{-1cm}
    \caption{Evolution of mass accretion rate at horizon with time for 3D simulations. MAD model shows rapid ejections, while SANE model shows steady accretion. INT model shows intermediate variability in accretion rate. The vertical lines in blue, orange and green represent the time corresponding to the snapshots for MAD, INT and SANE respectively provided in Figures \ref{fig:density_contours_MAD_xz_3D}, \ref{fig:density_contours_MAD_xy_3D}, \ref{fig:density_contours_INT_3D} and \ref{fig:density_contours_SANE_3D}.}
    \label{fig:mdot_t_3D}
\end{minipage}
\hfill
\begin{minipage}{0.48\textwidth}
    \includegraphics[width=\textwidth]{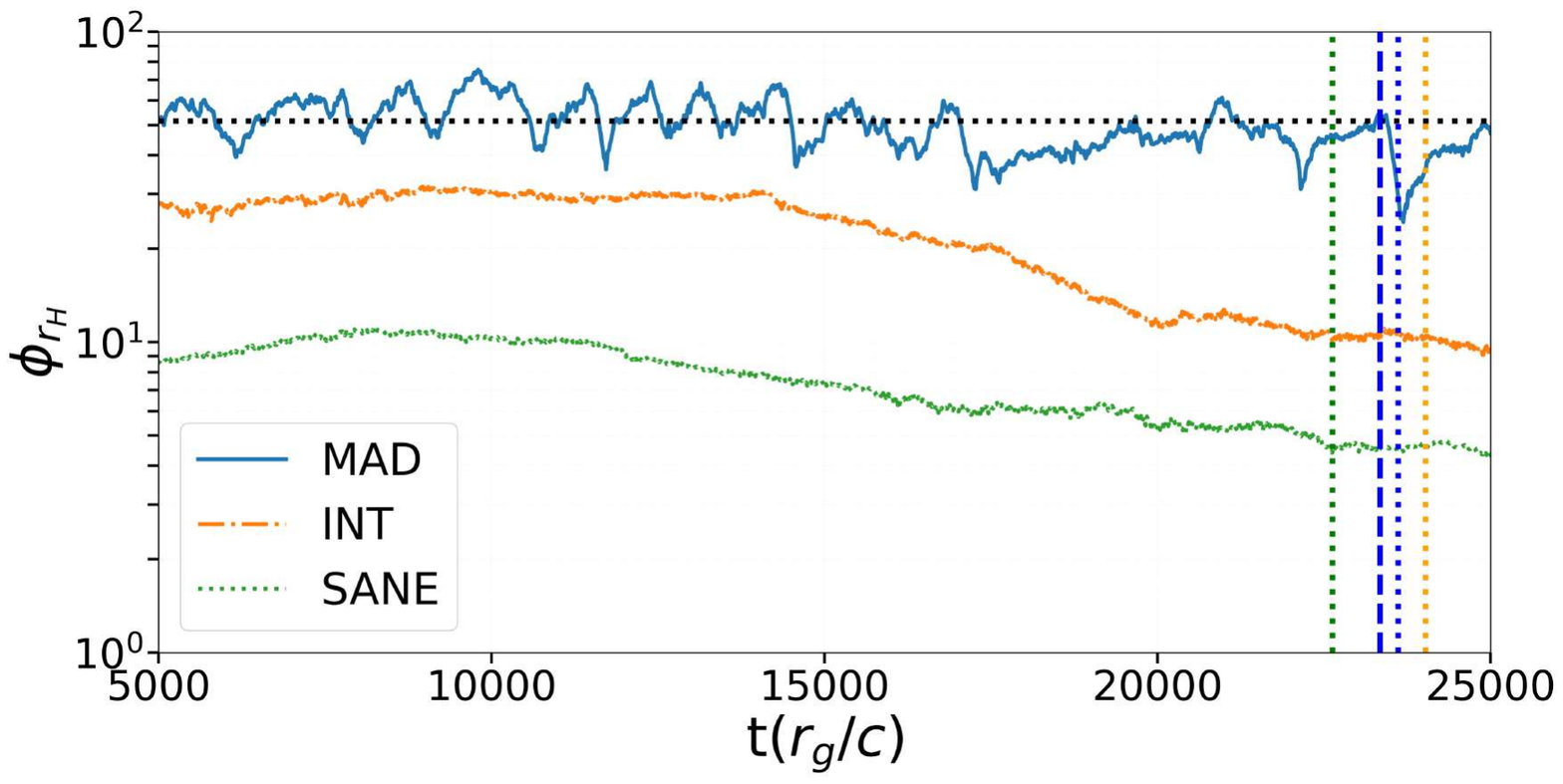}
    \vspace{-1cm}
        \caption{Normalized magnetic flux evolution with time for high resolution 3D simulations with the dotted horizontal line showing MAD saturation limit. MAD configuration reaches the saturation limit very easily and thus shows eruption of magnetic flux, SANE configuration always remains well below the saturation limit, whereas INT configuration shows an intermediate magnetic flux between MAD and SANE. Vertical lines are same as in Figure \ref{fig:mdot_t_3D}.}
    \label{fig:phi_t_3D}
\end{minipage}
\end{figure*}

\begin{figure}
    \includegraphics[width=0.48\textwidth]{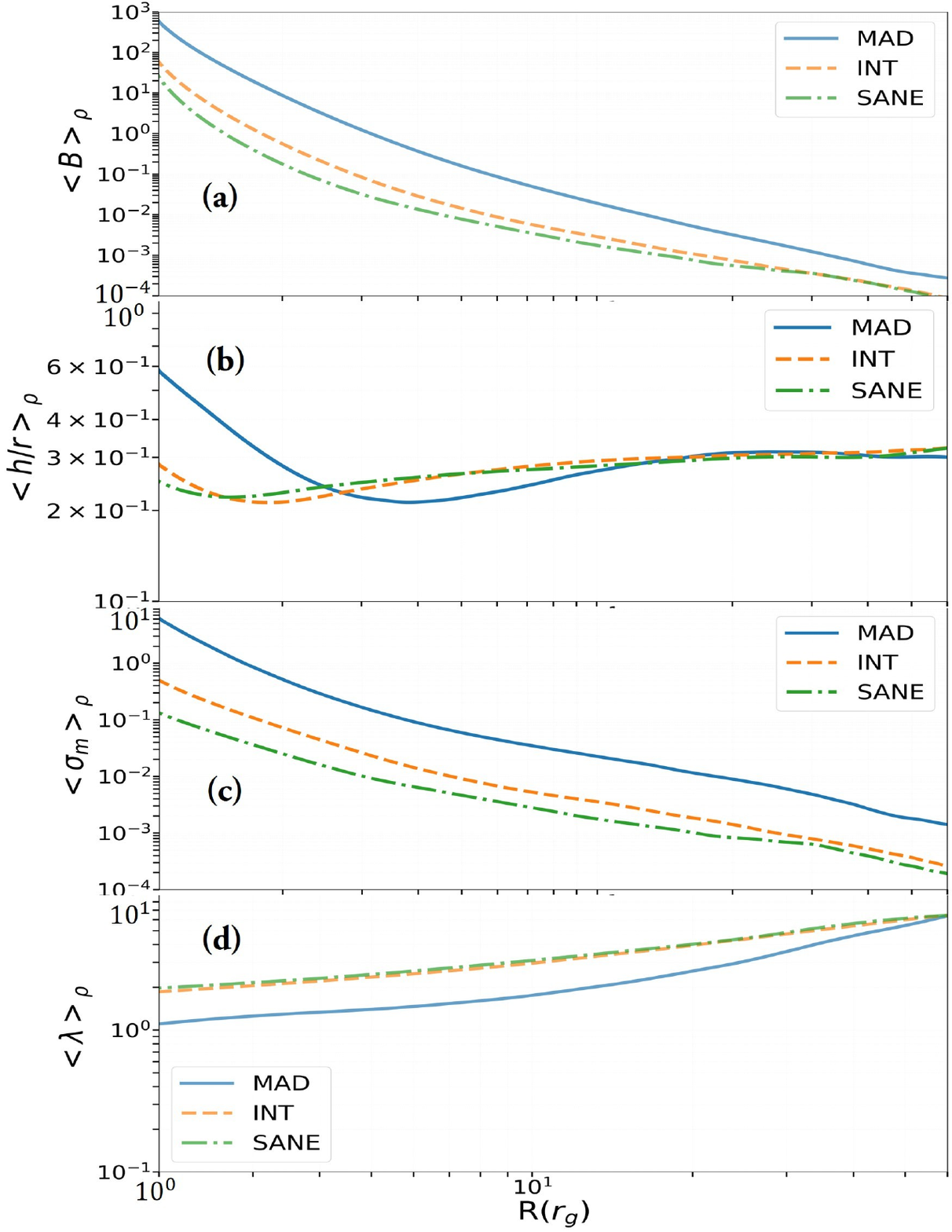}
    \caption{(a) Shell averaged total magnetic field in code units as a function of distance from black hole for 3D simulations. (b) Shell averaged disk aspect ratio as a function of distance from black hole for 3D simulations. (c) Magnetization parameter as a function of distance from the black hole for 3D simulations. (d) Shell averaged specific angular momentum as a function of distance from black hole for 3D simulations.}
    \label{fig:B_hr_sigma_lambda}
\end{figure}

\begin{figure}
    \includegraphics[width=0.48\textwidth]{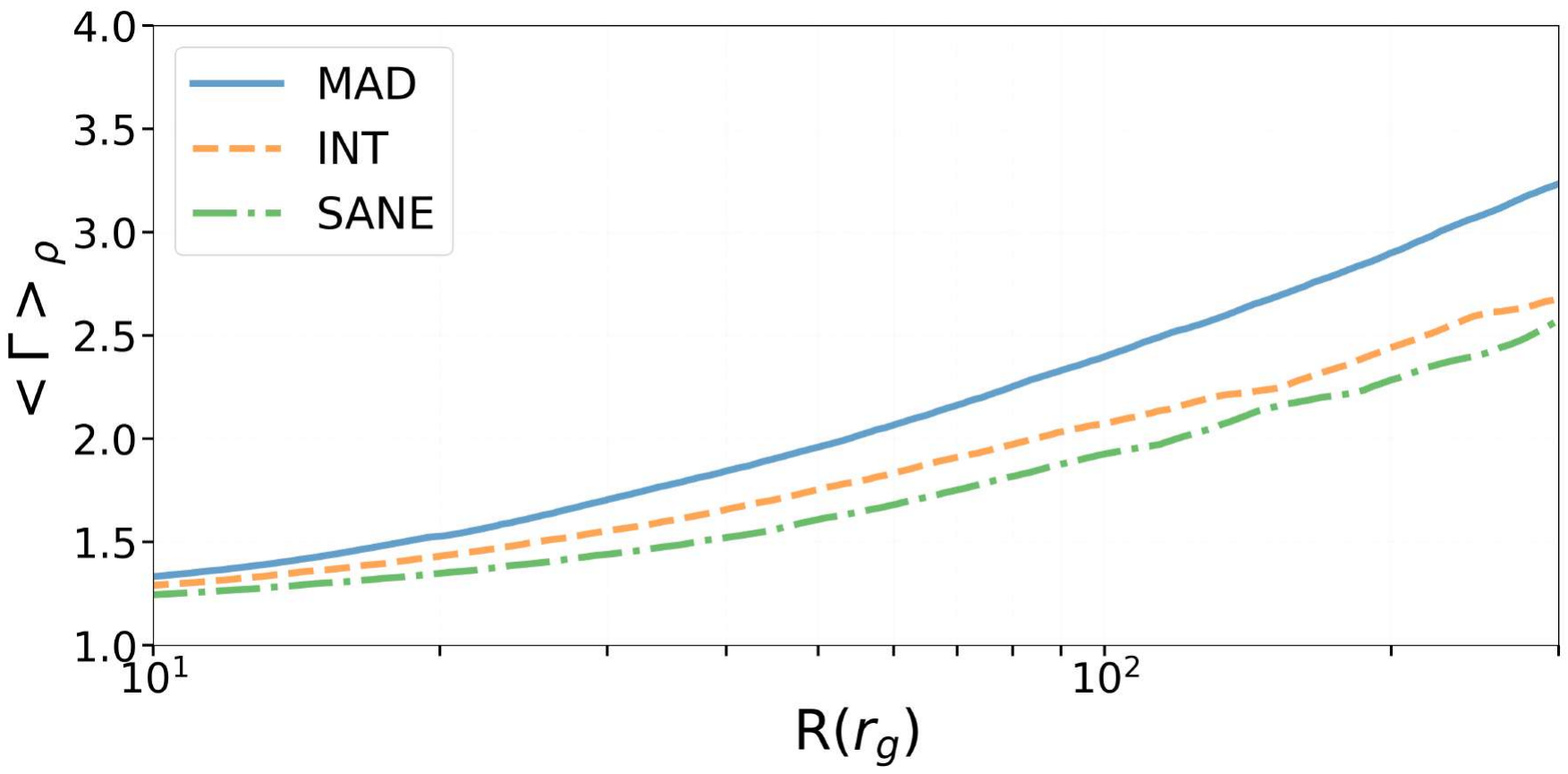}
    \vspace{-1cm}
    \caption{Lorentz factor as a function of distance from the black hole for 3D simulations.}
    \label{fig:lorentz_r_3D}
\end{figure}

\begin{table*}
\centering
\caption{Simulated properties of different 3D models and their characteristics}
\label{tab:parameters}
\scriptsize
\begin{tabular}{cccccccccccc}
\toprule
State & $|\langle v^r\rangle_\rho|$ & $\Gamma_{\rm max}$ & $\langle B_{\rm tot}\rangle_\rho$ & $\langle\lambda\rangle_\rho$ & $\langle\Omega/\Omega_K\rangle_\rho$ & $\langle\sigma_m\rangle_\rho$ & $\langle$plasma-$\beta\rangle_\rho$ & $\langle h/r\rangle_\rho$ & $\eta$ & $\dot{M}(r_H)$ & $\phi(r_H)$ \\
 & (at $r_H$) &  & (at $r_H$) & (at $r_H$) &  & (at $r_H$) & (at $r_H$) &  &  & variability & variability \\
\midrule
MAD & 0.10 & 3.23 & 1.06 & 0.60 & 0.55 & 5.89 & 0.12 & 0.2--0.6 & 1--1.5 & 0.236 & 0.150 \\
INT & 0.05 & 2.68 & 0.69 & 1.59 & 0.92 & 0.48 & 1.16 & 0.2--0.3 & 0.3--0.5 & 0.217 & 0.077 \\
SANE & 0.03 & 2.57 & 0.42 & 1.77 & 0.95 & 0.13 & 3.24 & 0.2--0.3 & 0.1--0.2 & 0.180 & 0.077 \\
\bottomrule
\end{tabular}\\
\vspace{2mm}
{\small All quantities in this table are derived from 3D GRMHD simulations (time-averaged over 20,000--25,000$r_g/c$):
   $|\langle v^r\rangle_\rho|$: Radial velocity at the horizon in units of $c$,
   $\Gamma_{\rm max}$: Maximum Lorentz factor of accelerated jets,
   $\langle B_{\rm tot}\rangle_\rho$: Total magnetic field strength at the horizon in code units,
   $\langle\lambda\rangle_\rho$: Specific angular momentum at the horizon,
   $\langle\Omega/\Omega_K\rangle_\rho$: Ratio of angular velocity to Keplerian angular velocity at outer radius,
   $\langle\sigma_m\rangle_\rho$: Magnetization parameter at the horizon,
   $\langle$plasma-$\beta\rangle_\rho$: Ratio of gas pressure to magnetic pressure,
   $\eta$: Outflow efficiency of the jet production or energy extraction,
   $\dot{M}_{r_H}$ variability: Mass accretion rate variability (computed as $\sigma_{\dot{M}}/\mu_{\dot{M}}$) where $\sigma_{\dot{M}}$ is the standard deviation and $\mu_{\dot{M}}$ is the mean of the $\dot{M}_{r_H}$ time series,
   $\phi_{r_H}$ variability: Magnetic flux variability (computed as $\sigma_\phi/\mu_\phi$) where $\sigma_\phi$ is the standard deviation and $\mu_\phi$ is the mean of the $\phi_{r_H}$ time series.
All quantities are shell-averaged over $\theta$ and $\phi$ directions.}
\end{table*}

\subsubsection{MAD State}

The MAD configuration (Figures~\ref{fig:density_contours_MAD_xz_3D}, \ref{fig:density_contours_MAD_xy_3D}) is characterized by rapid ejections evidenced by high variability in mass accretion (see Figure~\ref{fig:mdot_t_3D}). The normalized magnetic flux, $\phi$, exceeds 50 (see Figure~\ref{fig:phi_t_3D}), indicating significant magnetic field accumulation near the black hole \citep{Narayan2003}. This state exhibits strong magnetic fields (plasma-$\beta$ $\sim 0.1$, Figure \ref{fig:rho_pbeta_vr_om}b; $B>1$, Figure \ref{fig:B_hr_sigma_lambda}a; see Table~\ref{tab:parameters}) near the horizon, while maintaining a geometrically thick disk ($<h/r>_\rho\geq0.6$, see Figure~\ref{fig:B_hr_sigma_lambda}b, where $h$ is the half thickness of the disk and $r$ is the radius from black hole) supported by strong magnetic pressure ($<$plasma-$\beta>_\rho\ll1$; see Table~\ref{tab:parameters}).
The rotation profile in MAD shows significant sub-Keplerian motion ($<\Omega/\Omega_K>_\rho\sim0.5$, Figure \ref{fig:rho_pbeta_vr_om}d, where $\Omega$ is the angular velocity and $\Omega_K$ is the Keplerian angular velocity; see Table~\ref{tab:parameters}) resulting from strong magnetic support against gravity in the inner regions of the disk \citep{Chatterjee_2022} and thermal and centrifugal support together in the outer regions \citep{Begelman2022v2,Chatterjee_2022}. This sub-Keplerian nature comes with large shell averaged radial velocities ($|<v^r>_\rho|\sim0.1c$ at $r_H$, Figure \ref{fig:rho_pbeta_vr_om}c) and high velocity jets reaching relativistic speeds ($\Gamma_{max}\sim3$-$4$, $v_{jet}\sim c$, Figure \ref{fig:lorentz_r_3D};  see Table \ref{tab:parameters}) in the polar regions, where high magnetic pressure gradients accelerate the flow.  Note however that jet velocities in GRMHD simulations may be affected by artificial mass loading, though jet powers remain reliable. The high magnetization parameter ($<\sigma_m>_\rho\sim5$-$6$, Figure \ref{fig:B_hr_sigma_lambda}c; see Table \ref{tab:parameters}) provides natural channels for strong collimated jets. The jets are followed by a dip due to dragging of magnetic field lines by accretion leading to the formation of strong barrier which pushes the matter out, which are then collimated by strong magnetic fields. This organized flow structure leads to enhanced angular momentum transport, thus leading to a lower value of angular momentum at the horizon compared to other states as shown in Figure \ref{fig:B_hr_sigma_lambda}d and Table \ref{tab:parameters}. This angular momentum transport is primarily driven by magnetic stresses rather than hydrodynamic processes as previously shown by \citet{Chatterjee_2022}. The velocity field also shows clear evidence for magnetic barriers (Figures \ref{fig:density_contours_MAD_xz_3D}, \ref{fig:density_contours_MAD_xy_3D}) that periodically form and disrupt, creating alternating patterns of rapid outflow and infall  that characterize the MAD state.
The strongly sub-Keplerian inner flow creates conditions for magnetic dissipation-powered hard X-ray emission, while ordered fields maintain steady radio jets. While we cannot directly compute spectra, the combination of strong magnetic support, sub-Keplerian flow, and relativistic polar jets provides conditions where magnetic processes likely dominate the emission, naturally explaining the observed hard spectrum.

\subsubsection{INT State}

The INT configuration (Figure~\ref{fig:density_contours_INT_3D}), bridges MAD and SANE states with moderate magnetic flux ($\phi\sim20$-$40$, Figure \ref{fig:phi_t_3D}) and intermediate disk thickness ($<h/r>_\rho\sim0.3$, Figure \ref{fig:B_hr_sigma_lambda}b). The flow dynamics show transitional characteristics with moderately sub-Keplerian rotation ($<\Omega/\Omega_K>_\rho\sim0.92$, Figure \ref{fig:rho_pbeta_vr_om}d; see Table \ref{tab:parameters}) and balanced pressure support ($<$plasma-$\beta>_\rho\sim1$, Figure \ref{fig:rho_pbeta_vr_om}b; see Table \ref{tab:parameters}). Shell-averaged radial velocities reach $|<v^r>_\rho| \sim 0.08c$ at $r_H$ (Figure \ref{fig:rho_pbeta_vr_om}c; see Table \ref{tab:parameters}), while the 2D velocity field (see Figure \ref{fig:velocity_fields} for details) shows complex flow patterns with both inflow and outflow components. The velocity structure reveals less collimated outflows compared to MAD states, with maximum jet velocities $v_{jet} \sim 0.85c$ ($<\Gamma>_\rho \sim 2-3$, Figure \ref{fig:lorentz_r_3D}; see Table \ref{tab:parameters}) in the polar regions. Due to moderate magnetization ($<\sigma_m>_\rho \sim 0.5$, Figure \ref{fig:B_hr_sigma_lambda}c; see Table \ref{tab:parameters}), this state shows a balance between the presence of both jets and winds. This mixed morphology produces moderate angular momentum transport (Figure \ref{fig:B_hr_sigma_lambda}d, see Table \ref{tab:parameters} to get an idea of moderate values), with contributions from both magnetic and hydrodynamic processes. The INT configuration represents inherently unstable equilibria between MAD and SANE states. Our simulations demonstrate that this can occur when moderate magnetic fields ($<$plasma-$\beta>_\rho \sim 1$) are able to maintain corona-like conditions while undergoing continuous reorganization, as evident by the complex velocity field patterns.

\subsubsection{SANE State}

The SANE configuration (Figure~\ref{fig:density_contours_SANE_3D}) maintains consistently low magnetic flux ($\phi<20$, Figure \ref{fig:phi_t_3D}) with weak magnetic pressure ($<$plasma-$\beta>_\rho\gg1$, Figure \ref{fig:rho_pbeta_vr_om}b; see Table \ref{tab:parameters}). The flow dynamics reveal a structure dominated by centrifugal and gravitational forces rather than magnetic effects, with nearly Keplerian rotation ($\Omega/\Omega_K\sim 0.95$, Figure \ref{fig:rho_pbeta_vr_om}d; see Table \ref{tab:parameters}) and lower radial velocities ($|<v^r>_\rho|\sim0.02c$ at $r_H$, Figure \ref{fig:rho_pbeta_vr_om}c; see Table \ref{tab:parameters}) with evidence of turbulent eddies in the disk region with maximum jet velocities $v_{jet} \sim 0.1c$. This turbulent structure, driven by MRI rather than large-scale magnetic fields, results in less efficient angular momentum transport (Figure \ref{fig:B_hr_sigma_lambda}d; see Table \ref{tab:parameters}). Hence, being dominated by angular motion, the SANE configuration results in disks that are relatively thinner ($<h/r>_\rho \sim 0.2-0.3$, Figure~\ref{fig:B_hr_sigma_lambda}b) compared to MAD and INT states which are more quasi-spherical with stronger radial velocities. Also, low magnetization ($<\sigma_m>_\rho \sim 0.1$, Figure \ref{fig:B_hr_sigma_lambda}c; see Table \ref{tab:parameters}) signifies domination of disk winds instead of strong collimated jets. The absence of strong jets in SANE matches the underlying simulated velocity field (Figure \ref{fig:velocity_fields}) showing primarily turbulent motions rather than collimated jets.

\subsection{Outflow Properties}

The outflow power is calculated from the efficiency, $\eta$ (see Equation \ref{eqn:eff_tot} in Appendix), based on  $M$ and $\dot{M}$ of black hole, which is given by
\begin{equation}
    \label{eqn:power_formula}
    P=\eta \dot{M} c^2,
\end{equation}
where $\dot{M}=\dot{m}\dot{M}_{Edd}$ and Eddington mass accretion rate is given by 
 $\dot{M}_{Edd}=L_{Edd}/(\eta_{rad}c^2)=1.4\times10^{18}M/M_\odot$ where $\eta_{rad}=0.1$ is the radiative efficiency showing the amount of energy being radiated.
The efficiency parameter $\eta$ directly determines the jet power for a given accretion rate. 

The outflow power follows the fundamental scaling relation
\begin{equation}
   P_{out} \approx \eta\dot{M}c^2 \approx 10^{39}  \left(\frac{\eta}{0.1}\right)\left(\frac{M_{BH}}{10M_{\odot}}\right)\left(\frac{\dot{M}}{\dot{M}_{Edd}}\right) \text{ erg s}^{-1},
\end{equation}
where $\eta$ depends on magnetic configuration: $\eta \approx 1.0$-$1.5$ for MAD, $0.3$-$0.5$ for INT, and $0.1$-$0.2$ for SANE states (see Table \ref{tab:parameters}). This scaling, supported by observations across the mass range \citep{Webb2014,Miller2020}, demonstrates how similar accretion physics operates from stellar-mass XRBs to intermediate mass black holes.

\section{Comparison between 3D and 2D results}\label{sec:comparison}

\begin{table*}
\centering
\caption{Simulated properties of different models and their characteristics from high resolution 2D simulations}
\label{tab:comparison}
\begin{tabular}{ccccccccccc}
\toprule
State & $|\langle v^r\rangle_\rho|$ & $\Gamma_{\rm max}$ & $\langle B_{\rm tot}\rangle_\rho$ & $\langle\lambda\rangle_\rho$ & $\langle\sigma_m\rangle_\rho$ & $\langle$plasma-$\beta\rangle_\rho$ & $\langle h/r\rangle_\rho$ & $\eta$ & $\dot{M}(r_H)$ & $\phi(r_H)$ \\
 & (at $r_H$) & (at $r_H$) & (at $r_H$) &  & (at $r_H$) & (at $r_H$) &  &  & variability & variability \\
\midrule
MAD & 0.08 & 3.55 & 6.81 & 0.18 & 32.63 & 0.04 & 0.4--0.8 & 1--10 & 1.632 & 0.133 \\
INT & 0.08 & 2.12 & 6.06 & 0.93 & 25.28 & 0.11 & 0.2--0.4 & 1--2 & 1.066 & 0.245 \\
SANE & 0.02 & 1.84 & 4.35 & 1.85 & 4.04 & 1.11 & 0.1--0.2 & 0.7--1 & 0.227 & 0.108 \\
\bottomrule
\end{tabular}\\
\vspace{2mm}
{\small These quantities can be compared with those from 3D simulations in Table~\ref{tab:parameters}.}
\end{table*}

Our comparison between 2D and 3D simulations demonstrates that while there are some quantitative differences, the qualitative features and primary conclusions about accretion states remain robust. Analysis of mass accretion rates (as given in Table ~\ref{tab:comparison}) shows that both 2D and 3D simulations capture the key distinguishing characteristics: MAD states  exhibit high variability with quasi-periodic eruptions, INT states show moderate variability, and SANE states maintain steady accretion (see Figure \ref{fig:mdot_t_3D}, Table \ref{tab:parameters}).

The normalized magnetic flux evolution (see Table ~\ref{tab:comparison}) confirms that the threshold behavior identifying different accretion states persists in 2D. The MAD configuration readily achieves saturated flux with $\phi > 50$, while SANE remains consistently below the saturation threshold. 

Geometric properties like disk scale height (as listed in Table~\ref{tab:comparison}) show remarkable consistency between 2D and 3D, with $h/r$ profiles differing by less than 15\% across all configurations. The  plasma-$\beta$ distributions (also given in Table ~\ref{tab:comparison}) maintain the same hierarchical ordering in both 2D and 3D, though 3D cases show higher values with differences of more than an order of magnitude due to enhanced magnetic field dissipation through the additional degree of freedom. For the same reason, the magnetisation is also significantly higher in 2D simulations.

The key distinction between 2D and 3D simulations arises from the stronger accumulation of magnetic flux in 2D. This leads to a higher magnetic field strength at the horizon, which in turn suppresses accretion and reduces the radial infall velocity. Moreover, the MAD and INT states appear to be more effective at transporting angular momentum outward in 2D, resulting in lower angular momentum at the horizon. Together, these effects contribute to the enhanced variability observed in both the accretion rate and magnetic flux in 2D simulations.

\begin{table*}
\centering
\caption{Properties of different accretion states}
\label{tab:states}
\begin{tabular}{lccc}
\toprule
Property & MAD State & INT State & SANE State \\
\midrule
Magnetic field Strength & Strong (plasma-$\beta \ll 1$) & Moderate (plasma-$\beta \sim 1$) & Weak (plasma-$\beta \gg 1$) \\
\midrule
Mass Accretion & Highly Variable & Moderately Variable & Steady \\
 & Episodic ejections & Periodic Variations & Low Variability \\
\midrule
Outflow Properties & Strong Jets & Mixed Outflows & Weak winds \\
 & High Collimation & Moderate Collimation & Uncollimated \\
\midrule
Outflow Efficiency & $>100\%$ & $30-50\%$ & $<20\%$ \\
\midrule
Disk Structure & Geometrically thicker & Intermediate thickness & Geometrically thinner \\
 & Magnetically Supported & Partially Compressed & Centrifugally Supported \\
\midrule
Angular Momentum & Magnetic Domination & Mixed Transport & MRI Dominated \\
Transport & Strong Magnetic Stress & Moderate Stress & Weak Magnetic Stress \\
\bottomrule
\end{tabular}
\end{table*}

\section{Conclusions} \label{sec:conclusions}

Our systematic GRMHD simulations of accretion flows around rapidly spinning black holes (with $a=0.998$) reveal three distinct accretion states that is useful to explain the complex phenomenology observed across the black hole mass spectrum \citep{Narayan2022}. As summarized in Table \ref{tab:states}, these states are characterized by fundamentally different magnetic field configurations that drive distinctive outflow properties and variability patterns:

1. The MAD state, achieved with strong initial magnetic fields (plasma-$\beta<< 1$), exhibits powerful jets (power $\sim 10^{39}$ erg s$^{-1}$), highly variable accretion, and significant sub-Keplerian motion ($\Omega/\Omega_K \sim 0.2$-$0.4$). 

2. The INT state bridges the gap between MAD and SANE regimes, with moderate magnetic support (plasma-$\beta \sim 1$) producing mixed outflow morphologies and complex variability.

3. The SANE state, characterized by weak magnetic fields (plasma-$\beta >> 1$), shows steady accretion with primarily winds. 

Our results confirm the earlier theoretical $\theta$-averaged, axisymmetric outcome that different components of magnetic shear play important roles in controlling accretion dynamics. This is particularly for the large-scale strong magnetically arrested flows in both semi-analytic solutions \citep{Mukhopadhyay2015,Mondal2020} and simulations \citep{Chatterjee_2022}.

Our comparison between 2D and 3D simulations reveals that while quantitative details differ, the qualitative features distinguishing different accretion states remain robust. The primary difference appears in accretion efficiency, with 3D simulations requiring approximately five times higher accretion rates to achieve luminosities comparable to 2D cases. This difference arises from enhanced magnetic field dissipation \citep{Ripperda2022} and more complex flow structures in 3D, but importantly does not affect the fundamental relationships between magnetic field geometry, outflow properties, and variability characteristics.

A key finding of our work is the unification of seemingly disparate observational features through the framework of magnetic field evolution. The same underlying mechanisms - magnetic flux accumulation, eruption, and redistribution \citep{Tchekhovskoy2011} - seem to operate remarkably across   several orders of magnitude in black hole mass, explaining observations from stellar-mass XRBs to intermediate mass black holes. This suggests that magnetic field geometry, along with accretion rate (however below a critical value as per sub-Keplerian advective accretion flow), plays the decisive role in determining both spectral states and outflow properties. The success of this unified picture in explaining diverse phenomena, from relativistic jets to jet-disk connection to X-ray variabilities, demonstrates the fundamental importance of magnetic processes in governing black hole accretion, particularly for GRS~1915+105. Further, the success to possibly explain observations based on present non-radiative model suggests that radiation mechanism might have minimal (or no) impact to control disk dynamics for, at least, the systems we are addressing in the present work.

We suggest the following plausibilities, which however need to be confirmed by detailed radiation based simulations. 
The different temporal classes of GRS~1915+105 could be identified as MAD/INT/SANE: their disk and jet properties. The $\chi$ class might be MAD with the hardest spectra and steady powerful jets (of power $5$--$10\times10^{39}$ erg s$^{-1}$, considering an accretion rate of $\dot{M}\sim0.05\dot{M}_{Edd}$), while classes like $\nu$ and $\beta$ might correspond to INT states with variable outflows, and $\delta$ and $\lambda$ classes seem showing the SANE characteristics with only weak jets. The simulations reproduce observed jet velocities near the speed of light for MAD states, wind velocities $\sim1000$ km s$^{-1}$, and magnetic field strengths ranging from $10^8$ G at $r_H$ to $10^5$ G at larger distances. These different states are distinguished by their magnetization levels, with MAD having the highest ($\langle\sigma_m\rangle_\rho \sim 5.9$), INT intermediate ($\langle\sigma_m\rangle_\rho \sim 0.5$), and SANE the lowest ($\langle\sigma_m\rangle_\rho \sim 0.1$).

The simulation results could also extend to other XRBs across different mass scales. By attempting to map the results with Cyg~X-1, it could be identified as SANE-INT configuration, producing steady jet power $\sim10^{37}$ erg s$^{-1}$ for accretion rate of $\dot{M}\sim0.005\dot{M}_{Edd}$ with observed wind velocities matching simulated values $\sim0.02c$. HLX-1, an intermediate-mass black hole ($\sim20{,}000M_\odot$), seems demonstrating that these accretion mechanisms scale appropriately, producing jet powers of $10^{40}$--$10^{41}$ erg s$^{-1}$ with accretion rate of $\dot{M}\sim0.005\dot{M}_{Edd}$ in MAD states that match observed peak luminosities and jet velocities $\geq0.7c$. Apparent similarities between simulated and observed properties---including jet powers, wind velocities, magnetic field strengths---validate that magnetic field configuration plays a crucial role in determining the energetics and outflow characteristics of accreting black hole systems in hard states.  The high jet power in these systems can thus be explained even with very low accretion rates but very high efficiency due to large magnetization.
In future, a more rigorous radiative simulation only will establish the above mentioned plausibilities firmly.

It is important to note that our simulations are primarily applicable the hard and intermediate states of XRBs. The thermal or high/soft state, characterized by an optically thick, geometrically thin disk producing multicolor blackbody emission, requires efficient radiative cooling mechanisms absent in our non-radiative simulations. Some GRS~1915+105 classes with high diskbb contributions ($\mu$, $\phi$, and $\gamma$ classes with $\gtrsim 50\%$ diskbb contribution), therefore may not be corroborated with our simulations. Nevertheless, as those classes still maintain a considerable power-law component, we suggest that some of them may be the combination of SANE and standard Keplerian accretion models. These classes likely represent states where radiative cooling becomes important while corona/sub-Keplerian activity continues to produce the non-thermal emission. The full thermal state might emerge from significantly ``modified SANE" configurations if radiative processes were included, potentially representing a third major accretion state alongside the MAD and INT states focused on in this work.

While our simulations consider different initial conditions to produce distinct accretion states, they do not capture real-time transitions between states. However, we try to speculate the unified picture, at least partially, for how these transitions might occur in real systems. Our unified framework of three distinct accretion states suggests possible mechanisms involved with magnetic fields driving state transitions and hysteresis. 
 For the generation of hard/intermediate state, magnetic fields have been shown to accumulate near the black hole via advection in the disk from larger-scales \citep{Narayan2003} or via in-situ generation within the disk \citep{King2004}. The transition from hard to soft states is, however, an open question. While the supply of magnetic fields via advection is much less efficient when the accretion flow is geometrically thinner (hence they form near Keplerian angular momentum distribution), the mechanism behind the diffusion of magnetic fields from near the black hole \citep{Lubow1994} is unknown, as GRMHD simulations are yet unable to capture the relevant timescales for state transitions (which vary over days to weeks and even months). One possibility being however the supply of magnetic flux is sluggish, whatever be the reason. Magnetic flux eruptions, also seen in our MAD simulations, tend to redistribute the magnetic flux within the disk \citep{Tchekhovskoy2011}, with a tiny amount of flux dissipated away. This process could, in principle, account for the loss of magnetic flux over long timescales. However, no simulation has yet reliably demonstrated whether the rate of loss of magnetic flux is significant.

Nevertheless, while our simulations capture the essential physics of different accretion states and outflows/jets, several important limitations should be addressed in future work. First, the inclusion of radiative cooling and radiation pressure effects \citep{McKinney_2012, Chatterjee_2023} would provide more accurate modeling of spectral states and transition timescales. \citet{Liska2024}  performed radiative two temperature GRMHD simulations of truncated disks where they have shown how the SANE disk collapses to a standard thin disk when cooling effects are included, whereas the MAD disk remains advective. The thin disk part of the SANE disk could then be the source of soft photons which form the diskbb component of the softer spectral states of GRS~1915+105. Second, our treatment of magnetic field dissipation \citep{Chatterjee_2022} could be improved through higher-resolution studies of reconnection physics \citep{Ripperda2022, Salas2024}. Third, the impact of more realistic equation of state and detailed microphysics should be investigated. Future work should also explore the role of misaligned accretion flows and non-axisymmetric perturbations in driving state transitions.

Videos of our simulations showing the temporal evolutions of density, magnetic fields, and outflows in different accretion states are available at \url{http://www.youtube.com/@rohanraha2157}. These visualizations provide additional insight into the complex dynamics revealed by our calculations.

\section*{acknowledgments}
The authors thank the anonymous referee for providing valuable and profound comments. RR acknowledges the Prime Minister’s Research Fellowship (PMRF) scheme for providing fellowship. BM would like to acknowledge the project funded by SERB, India, with Ref. No. CRG/2022/003460, for partial support to this research. We thank Arif Babul (UVic), Ramesh Narayan (Harvard), Diego Altamirano (Southampton) and A. R. Rao (TIFR) for helpful discussions. KC would like to acknowledge Compute Canada (http://www.computecanada.ca) for computational resources used for this project.

\section*{Data Availability}

The data that support the findings of this article are based on rigorous computer simulations and are not publicly available upon publication because it is not technically feasible and/or the cost of preparing, depositing, and hosting the data would be prohibitive within the terms of this research project. The data are available from the authors upon reasonable request.



\bibliographystyle{mnras}
\bibliography{ref} 




\appendix

\section {Numerical Setup} \label{sec:setup}

We use the HARMPI/H-AMR code to solve the GRMHD equations for a radiatively inefficient magnetized accretion flow around a rotating black hole defined by Kerr metric in Kerr-Schild coordinates whose line element is given by
\begin{eqnarray}
&ds^2=-\left(1-\frac{2r}{\Sigma}\right)dt^2+\left(\frac{4r}{\Sigma}\right)dr dt+\left(1+\frac{2r}{\Sigma}\right)dr^2+\Sigma d\theta^2 \\ \nonumber
&+\sin^2\theta\left[\Sigma+a^2\left(1+\frac{2r}{\Sigma}\right)\sin^2\theta\right]d\phi^2\\ \nonumber
&-\left(\frac{4ar\sin^2\theta}{\Sigma}\right)d\phi dt -2a\left(1+\frac{2r}{\Sigma}\right)\sin^2\theta d\phi dr,
\end{eqnarray}
where $\Sigma\equiv r^2+a^2 \cos^2\theta$ and $g\equiv Det(g_{\mu\nu})=-\Sigma^2\sin^2\theta$.
Unless otherwise noted, we use units of $c=GM=1$, and we follow the notational conventions of \cite{Misner:1973prb}. The code employs a conserved scheme that maintains both the divergence-free constraint on magnetic fields and conservation of mass-energy-momentum to machine precision. The detailed equations of accretion flow, governed by the conservation of stress-energy tensor along with continuity equation and magnetic induction equation, are given in several papers, see, e.g., \citealt{Gammie2003}, \citealt{Liska_2022}.

Our numerical integrations are carried out in a modified Kerr-Schild coordinates $x_0$, $x_1$, $x_2$ and $x_3$, given by 
\begin{eqnarray}
    &x_0&=t\\ \nonumber
    &x_3&=\phi\\ \nonumber
    &r&=e^{x_1}\\ \nonumber
    &\theta&=\pi x_2+\frac{1}{2}(1-h)\sin(2\pi h x_2).
\end{eqnarray}

Our 2D and 3D simulation grids extend from $r = r_{in} = 0.87 r_H$ to $10^3r_g$, with $r_H$ being the radius of the event horizon and $r_g$ is the gravitational radius $r_g = GM_{BH}/c^2$, ensuring we capture both the dynamics near the black hole and large-scale magnetic field evolution. The grid resolution used in 2D is $N_r \times N_\theta \times N_\phi \equiv 256 \times 256 \times 1$ for low resolution and $N_r \times N_\theta \times N_\phi \equiv 448 \times 192 \times 1$ for high resolution. For 3D simulations, the grid resolution used is $N_r \times N_\theta \times N_\phi \equiv 448 \times 240 \times 192$. These resolutions are chosen to adequately resolve both the MRI that drives turbulent transport and the large-scale magnetic structures that develop in MAD states. We apply outflowing radial boundary conditions (BCs), transmissive polar BCs, and periodic BCs in the azimuthal direction.

Our simulation starts with a black hole surrounded by a standard Fishbone-Moncrief (FM) torus  \citep{1976ApJ...207..962F}. 
 We use an
ideal gas equation of state with the gas pressure $p_{gas} =(\tilde\Gamma-1)u_{gas}$, where $\tilde\Gamma = 13/9$ is the adiabatic index and $u_{gas}$ is the internal energy of the gas. 

We initialize our model with different magnetic field geometries as given in Appendix \ref{sec:simulations} (see Table \ref{tab:simulations}) \citep{McKinney_2012,Chatterjee_2022} to explore different pathways to magnetic flux accumulation. The magnetic field strength in the initial setup is normalized by
setting max($p_{gas}$)/max($p_{mag}$), where $p_{mag} = b^2/2$, is the
magnetic pressure.

To ensure our simulations adequately resolve the MRI, we calculate the quality factors $Q_{\text{MRI}}$ for each model following the methodology of \citet{Hawley2011} and \citet{McKinney_2012}. $Q_{\text{MRI}}$ represents the number of grid cells per fastest-growing MRI wavelength in each direction and is defined as

\begin{equation}
Q_{x,\text{MRI}} \equiv \frac{\lambda_{x,\text{MRI}}}{\Delta x},
\end{equation}
where $\lambda_{x,\text{MRI}}$ is the fastest-growing MRI wavelength in the $x$-direction, and $\Delta x$ is the grid spacing in that direction. The MRI wavelength is given by

\begin{equation}
\lambda_{x,\text{MRI}} \approx 2\pi \frac{|v_{x,A}|}{|\Omega_{\text{rot}}|},
\end{equation}
for $x = r, \theta, \phi$, where $|v_{x,A}| = \sqrt{b^x b_x / \epsilon}$ is the $x$-directed Alfvén speed, $\epsilon \equiv b^2 + \rho + u_{gas} + p_{gas}$, and $r\Omega_{\text{rot}} = v_{\text{rot}}$ where $v_{rot}$ is the rotational velocity of the disk.

\begin{table}
\centering
\caption{MRI Quality Factors averaged in the bulk of the disk between $r = 20 r_g$ and $r = 60 r_g$}
\label{tab:qmri}
\begin{tabular}{lccc}
\hline
Component & MAD & INT & SANE \\
\hline
$Q_{r,\text{MRI}}$ & 82.45 & 14.47 & 10.72 \\
$Q_{\theta,\text{MRI}}$ & 93.69 & 14.28 & 11.11 \\
$Q_{\phi,\text{MRI}}$ & 75.48 & 25.35 & 21.10 \\
\hline
\end{tabular}
\end{table}

Table~\ref{tab:qmri} presents the averaged $Q_{\text{MRI}}$ values for our three primary accretion states between $r = 20 r_g$ and $r = 60r_g$. 
The MAD model exhibits exceptional resolution with values close to 100  for all components, while INT and SANE models maintain $Q_{\text{MRI}} > 10$ throughout the domain. The value of $Q_{\phi,\text{MRI}}$ shows the highest values, especially for INT and SANE states, consistent with our grid design that emphasizes resolving non-axisymmetric modes. All models significantly exceed the minimum threshold of $Q_{\text{MRI}} > 6$ required for properly resolving MRI-driven turbulence \citep{Porth2019, Sano2004}. The particularly high $Q_{\text{MRI}}$ values in MAD models reflect the strong magnetic field amplification characteristic of this configuration.

\section {Calculations and additional 2D/3D simulation results} \label{sec:2dresults}
For flux densities $F_d$, the flux integral is calculated as 
\begin{equation}
F(r)=\int \sqrt{-g}F_d d\theta d\phi.
\end{equation}
For weight $w$, the average of Q is 
\begin{equation}
    Q_w(r)=<Q>_w=\frac{\int \sqrt{-g} w Q d\theta d\phi}{\int \sqrt{-g} w d\theta d\phi}.
\end{equation}
All physical quantities like velocity and magnetic field are weighted by density.

All quantities are computed from the results produced in every $\sim10r_g/c$.
For all quantities $Q$, averages over time are performed from $t=15,000 r_g/c$ to $t=20,000 r_g/c$ for low resolution 2D models, from $t=25,000 r_g/c$ to $t=30,000 r_g/c$ for high resolution 2D models and from $t=20,000 r_g/c$ to $t=25,000 r_g/c$ for 3D models.

The mass accretion rate as a function of radius ($r$) is calculated as follows:
\begin{equation}
    \dot{M}(r)=-\int\sqrt{-g} \rho u^r d\theta d\phi.
\end{equation}

To confirm that our simulations have reached inflow equilibrium, we present the radial profile of the mass accretion rate in Figure \ref{fig:mdot_all}. It demonstrates that the total $\dot{M}$ remains nearly constant with radius in the inner parts of the disk, verifying that the necessary equilibrium conditions are satisfied. For SANE state, this constancy with radius directly indicates a stable inflow equilibrium. However, for the MAD state, the inherent variability necessitates a time-averaged profile over multiple variability cycles, which similarly confirms a statistically stationary state in the mean, free from domination by transient effects.

The dimensionless radial magnetic flux threading a black hole  normalized by mass accretion rate is given according to \citet{Tchekhovskoy2011} as
\begin{equation}
    \phi=\frac{\sqrt{4\pi}}{2\sqrt{\dot{M}}}\int \sqrt{-g}|B^r| d\theta d\phi,
    \label{eqn:mag_flux}
\end{equation}
where $B^r$ is the radial component of the magnetic field. MAD systems are characterized by a flow with $\phi$ values saturated at $~30-50$ \citep{Narayan2022}.

The approximation given by \citet{Narayan2022} of magnetic flux for MAD systems is
\begin{equation}
    \phi_{MAD}(a)=52.6+34a-14.9a^2-20.2a^3.
    \label{eqn:phi_mad_approx}
\end{equation}

The rate of mass outflux is calculated as:
\begin{equation}
\dot{M}_{\text{out}}(r) = \int \rho \max(0, u^r) \sqrt{-g} \, d\theta d\phi,
\label{eq:mdot_out}
\end{equation}
where the $\max(0, u^r)$ function selects only outflowing material with positive radial velocity.

The shell-integrated angular momentum flux is
\begin{equation}
\mathcal{J}^r_{\text{int}}=\int \sqrt{-g} T^r_\phi d\theta d\phi,
\label{eq:specific_j}
\end{equation}
and the specific angular momentum flux normalized by the mass accretion rate is given by:
\begin{equation}
\hat{\dot{J_r}}/\dot{M}_{5r_g} = \frac{\mathcal{J}^r_{\text{int}}\sqrt{g_{rr}}}{\dot{M}_{5r_g}},
\label{eq:specific_j}
\end{equation}
where hat represents the orthonormalized components like $\hat{Q_i}=Q^i\sqrt{g_{ii}}$ and $\dot{M}_{5r_g}$ is the mass accretion rate at $5r_g$.

The density ($\rho$) and velocity ($v^r$) are calculated by taking average over time  and then weighted by density, while integrating over a shell in $\theta$ and $\phi$ direction as follows:
\begin{equation}
    \rho_\rho=\frac{ \int\sqrt{-g}\rho^2 d\theta d\phi}{ \int\sqrt{-g} \rho d\theta d\phi},
\end{equation}

\begin{equation}
    v^r_\rho=\frac{ \int\sqrt{-g}\rho \frac{u^r}{u^t}d\theta d\phi}{ \int\sqrt{-g} \rho d\theta d\phi},
\end{equation}
The density-weighted disk averaged angular velocity is similarly calculated as 
\begin{equation}
    \Omega_\rho=\frac{ \int \sqrt{-g}\rho \frac{u^\phi}{u^t} dd\theta d\phi}{ \int\sqrt{-g} \rho d\theta d\phi},
\end{equation}
and the Keplerian angular momentum is given as $\Omega_K=1/r^{3/2}$.

The density-weighted disk averaged magnetic field is given by

\begin{equation}
    B_\rho=\frac{\int \sqrt{-g} \rho B d\theta d\phi}{ \int\sqrt{-g} \rho d\theta d\phi},
\end{equation}
where $B$ is the total magnetic field defined as $B=\sqrt{B_r^2+B_\theta^2+B_\phi^2}$.

The plasma-$\beta$ is defined as the ratio of gas pressure to the magnetic pressure, calculated as 
\begin{equation}
    \text{plasma}-\beta_\rho=\frac{ \int\sqrt{-g}\rho p_g d\theta d\phi}{ \int \sqrt{-g}\rho p_m d\theta d\phi},
\end{equation}
 where $p_m=b^2/2$ is the magnetic pressure and $p_g$  is the gas pressure.

We use the magnetization parameter to characterize the
ratio of the magnetic energy density to the rest mass energy
density, given by \citet{Narayan2022} 
\begin{equation}
    {\sigma_m}=b^2/\rho.
    \label{eqn:sigma}
\end{equation}

In order to calculate the jet velocity we use the Lorentz factor which is given by
\begin{equation}
    \Gamma_\rho=\frac{\int\sqrt{-g} (\sigma_m>1) \alpha u^t d\theta d\phi}{\int\sqrt{-g} (\mu>2) d\theta d\phi}
    \label{eqn:lorentz}
\end{equation}
where $\alpha=1/\sqrt{-g^{tt}}$.

\citet{Porth2019} defined the disk scale height to radius ratio $h/r$ as
\begin{equation}
    (\frac{h}{r})_\rho=\frac{ \int \sqrt{-g}\rho|\pi/2-\theta| d\theta d\phi}{ \int \sqrt{-g}\rho d\theta d\phi}.
    \label{eqn:aspect_ratio}
\end{equation}

The total radial energy flux is given by 
\begin{equation}
\label{eq:edot_tot}
    \dot{E}=- \int\sqrt{-g} T^r_t d\theta d\phi,
\end{equation}
where $T^{\mu\nu}=(\rho+u+p+b^2)u^\mu u^\nu +(p+\frac{1}{2}b^2) g^{\mu\nu}-b^\mu b^\nu$. The hydrodynamic and electromagnetic components of the radial energy flux are similarly given by 
\begin{eqnarray}
    \dot{E}_{HD}=- \int\sqrt{-g} T^r_t(HD) d\theta d\phi, \label{eq:edot_HD}\\
    \dot{E}_{EM}=- \int\sqrt{-g} T^r_t(EM) d\theta d\phi,  \label{eq:edot_EM}
\end{eqnarray}
where $T^{\mu\nu}(HD)=(\rho+u+p)u^\mu u^\nu +p g^{\mu\nu}$ and $T^{\mu\nu}(EM)=b^2 u^\mu u^\nu +\frac{1}{2} b^2 g^{\mu\nu}-b^\mu b^\nu$.

The net outflow efficiency is given by 
\begin{equation}
    \eta=\frac{\dot{E}-\dot{M}}{\dot{M}}=\frac{P_{out}}{\dot{M}},
    \label{eqn:eff_tot}
\end{equation}
which refers to the power of outflows that escape out to infinity normalized by the mass accretion rate.

The hydrodynamic and electromagnetic outflow efficiencies are similarly calculated as 
\begin{eqnarray}
    \eta_{HD}=\frac{\dot{E}_{HD}-\dot{M}}{\dot{M}}=\frac{P_{out}^{HD}}{\dot{M}} ,\label{eq:eff_HD}\\
    \eta_{EM}=\frac{\dot{E}_{EM}-\dot{M}}{\dot{M}}=\frac{P_{out}^{EM}}{\dot{M}}.  \label{eq:eff_EM}
\end{eqnarray}

Positive values of efficiency correspond to the extraction of energy from the system in the form of jets or outflows.

\section{Simulation Survey and Exploration of 2D Low Resolution Models} \label{sec:simulations}

 Our GRMHD simulations explore how magnetic field geometry and plasma-$\beta$ influence accretion dynamics around rapidly spinning black holes (with spin parameter $a = 0.998$). Using the HARMPI code, we study transitions between different accretion states that may explain the complex variability and transitions observed in XRBs. However, our GRMHD setup is by definition advective and sub-Keplerian, hence it can not capture the soft accretion state which is the Keplerian accretion disk.

 \begin{table*}
\centering
\caption{Summary of 2D and 3D GRMHD simulations for $a = 0.998$ with various initial plasma-$\beta$ and magnetic field geometries}
\begin{tabular}{llccll}
\hline
Model & Field type & plasma-$\beta_0$ & Resolution & Magnetic potential & Geometry description \\
\hline
P1B01-2DLR & Poloidal1 & 0.1 & $256\times256\times1$ & $A_\phi = \max(q,0)$ & Creates a simple poloidal field confined to \\
P1B01-2DHR & & 0.1 & $448\times192\times1$ & $q = \rho/\rho_{max} - 0.2$ & high-density regions with field lines. \\
P1B1-2DLR & & 1 & $256\times256\times1$ & & This setup concentrates magnetic flux in \\
P1B10-2DLR & & 10 & $256\times256\times1$ & & the densest parts of the accretion flow, \\
P1B100-2DLR & & 100 & $256\times256\times1$ & & potentially leading to magnetically driven \\
P1B100-2DHR & & 100 & $448\times192\times1$ & & outflows from these regions.  \\
P1B100-3D & & 100 & $448\times240\times192$ & &  \\
P1B1000-2DLR & & 1000 & $256\times256\times1$ & &  \\
\hline
P2B01-2DLR & Poloidal2 & 0.1 & $256\times256\times1$ & $A_\phi = \max(q,0)$ & More complex configuration with additional \\
P2B1-2DLR & & 1 & $256\times256\times1$ & $q = \rho/\rho_{max}(r/r_{in})^3$ & radial and angular dependencies. Creates a \\
P2B10-2DLR & & 10 & $256\times256\times1$ & $\sin^3\theta e^{-r/400} - 0.2$ & stronger field near the equatorial plane that \\
P2B100-2DLR & & 100 & $256\times256\times1$ & & falls off with radius, promoting vertical \\
P2B100-2DHR & & 100 & $448\times192$ & & field structures and stronger magnetic \\
P2B100-3D & & 100 & $448\times240\times192$ & & activity in the inner disk regions. This \\
P2B1000-2DLR & & 1000 & $256\times256\times1$ & & geometry enables rapid flux accumulation. \\
\hline
PLB01-2DLR & Poloidal & 0.1 & $256\times256\times1$ & $A_\phi = \max(q^2r^3,0)$ & Creates a magnetic field that scales with both \\
PLB1-2DLR & Long loop & 1 & $256\times256\times1$ & $q = \rho/\rho_{max} - 0.2$ & density and radius. Field is strongest in \\
PLB10-2DLR & & 10 & $256\times256\times1$ & & high-density regions and increases rapidly \\
PLB100-2DLR & & 100 & $256\times256\times1$ & & with distance. The quadratic dependence on \\
PLB100-2DHR & & 100 & $448\times192\times1$ & & $q$ produces stronger contrast between high \\
PLB100-3D & & 100 & $448\times240\times192$ & & and low density regions than Poloidal1. The \\
PLB1000-2DLR & & 1000 & $256\times256\times1$ & & $r^3$ factor drives strong outer activity. \\
\hline
\end{tabular}
\label{tab:simulations}
\end{table*}

We explore how different initializations lead to distinct evolutionary paths (listed in Table \ref{tab:simulations}) to explore the full range from MAD to SANE states. Our simulations explore how different initial magnetic field configurations and plasma-$\beta$ values lead to 
distinct evolutionary paths. We categorize our simulations into 3 groups: P1, P2 and PL, depending on the magnetic field geometry. P1 group creates a simple poloidal field in high-density regions, concentrating magnetic flux. P2 group has a more complex setup, resulting in a stronger field near the equatorial plane that can promote vertical field structures. PL group creates a magnetic field that scales with both density and radius, with the field being strongest in high-density regions. For each of the three groups described above, we have run simulations with initial plasma-$\beta$ values starting from 0.1 to 1000. The models are named based on their group followed by the initial plasma-$\beta$ value (for example, P1B100 model means initial plasma-$\beta=100$ and the magnetic field geometry is of P1 type). The time evolution of various parameters in our simulations, like accretion rate and magnetic flux, reveals the complex dynamics of accretion flows around rapidly spinning black holes and provides key diagnostics for identifying different accretion states.

In this section, we do our preliminary analysis with low resolution 2D simulations of all the models. Based on these 2D results, we select P2B100, PLB100, and P1B100 as representative models for detailed 3D analysis (presented in Section \ref{sec:results} of the main text). Finally, we will compare high resolution 2D and 3D simulations and discuss the differences.

All quantities are computed from the results produced in every $\sim10r_g/c$ where the gravitational radius $r_g=GM/c^2$.
Time averages of quantities $Q$ are represented as $<Q>$, averaged over time  from $t=15,000 r_g/c$ to $t=20,000 r_g/c$ for low resolution 2D models, from $t=25,000 r_g/c$ to $t=30,000 r_g/c$ for high resolution 2D models, and from $t=20,000 r_g/c$ to $t=25,000 r_g/c$ for 3D models. This temporal window represents the quasi-steady state phase of the simulations after initial transients have died away, with steady-state conditions achieved out to $r\approx60r_g$ as verified by the radial mass accretion rate profile (see Appendix \ref{sec:2dresults}). Density weighted time averaged quantities are represented as $<Q>_\rho$. All radial profiles are shell averaged in $\theta$ and $\phi$ directions.

\begin{figure*}
  \centering
\vspace{-1cm}
  \includegraphics[width=1.05\textwidth]{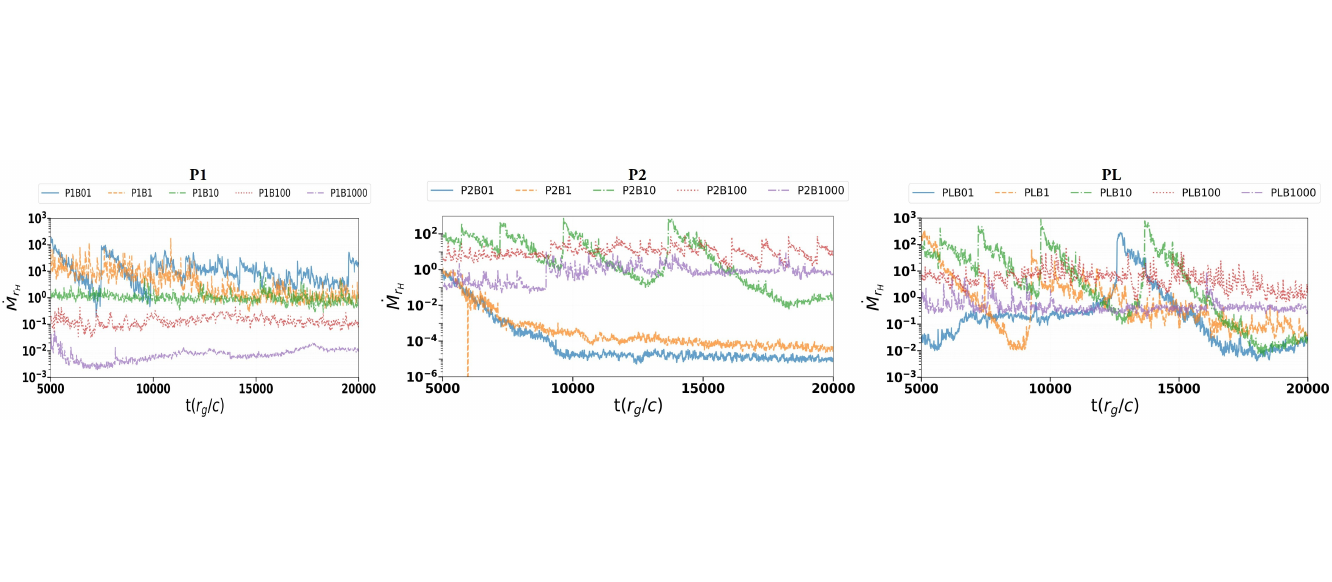}
\vspace{-2.5cm}
  \caption{Evolution of dimensionless mass accretion rate with time for P1, P2 and PL configurations for low resolution 2D models.}
  \label{fig:mdot_t_LR}
\end{figure*}

Figure \ref{fig:mdot_t_LR} shows the evolution of mass accretion rate at horizon ($r_H$) in code units for the P1, P2, and PL configurations. All configurations exhibit significant variability in the accretion rate, but with distinct characteristics. For the P1 configurations, we observe that lower plasma-$\beta$ values lead to higher average accretion rates with more pronounced variability. This is consistent with the idea that stronger magnetic fields can drive more efficient angular momentum transport through magnetic stresses \citep{McKinney_2012, Cao2013, Mukhopadhyay2015, Begelman2022v2}. The highest accretion rates are seen for plasma-$\beta = 0.1$ and $1$, indicating that these configurations achieve the most effective mass inflow through the combined effects of magnetic pressure gradients (shear) and flux eruption cycles. The variability in accretion rates for plasma-$\beta=0.1,1$ and $10$ cases is characteristic of MADs which form magnetic barriers to prevent matter from flowing in. However, we observe steady variation in cases of plasma-$\beta=100$ and $1000$, characteristic of SANE.

The P2 configurations show a different trend, with the highest accretion rates occurring at intermediate plasma-$\beta$ values ($10-100$), as shown in Figure \ref{fig:mdot_t_LR}. This non-monotonic behavior suggests a complex interplay between magnetic field strength and geometry in determining the accretion efficiency. For the very low plasma-$\beta$ cases, the matter is kicked out due to the formation of strong magnetic barriers by the accumulation of extreme magnetic flux. Since matter is not present, the accretion rate is very low for those cases. It also shows that, unlike the P1 configurations, the P2 configurations can accumulate magnetic flux very fast, which will be seen in detail in the next subsection. The plasma-$\beta=100$ case shows high variability similar to MAD state and the plasma-$\beta=1000$ case shows steady accretion like SANE state.

The PL configuration has the highest accretion rates occurring at high plasma-$\beta$ values ($100-1000$), as shown in Figure \ref{fig:mdot_t_LR}. In this case, the matter is also kicked out for the plasma-$\beta=0.1$ and $10$ cases, as seen by the decaying mass accretion rate with time, like low plasma-$\beta$ cases of P2. For plasma-$\beta=100$ and $1000$ cases, behaviour is similar to that in P2.

\begin{figure*}
  \centering
\vspace{-3cm}
  \includegraphics[width=1.05\textwidth]{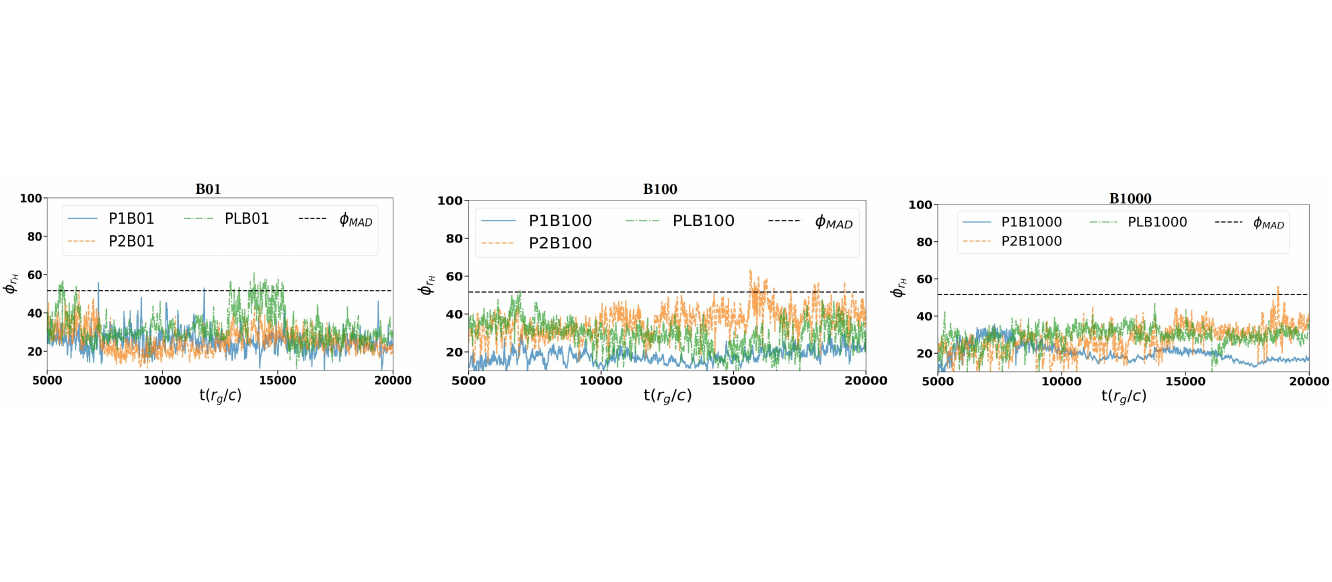}
\vspace{-2cm}
  \caption{Evolution of normalized magnetic flux at horizon with time for initial plasma-$\beta=0.1,100$ and $1000$ configurations for low resolution 2D models with the horizontal dotted line representing magnetic flux saturation level.}
  \label{fig:phi_t_LR}
\end{figure*}

The evolution of normalized magnetic flux threading the black hole, shown in Figure \ref{fig:phi_t_LR} for plasma-$\beta=0.1,100$ and $1000$, provides insight into the development of magnetically arrested states. The horizontal dotted line in Figure \ref{fig:phi_t_LR} represents the saturated magnetic flux values (see equation \ref{eqn:phi_mad_approx} in Appendix). When the system has accumulated enough magnetic flux above the saturated values as seen in the figures, flux eruption events happen which remove the flux from the system and there is a simultaneous drop in inflow mass accretion rate as was earlier seen in Figure \ref{fig:mdot_t_LR}.

For the P1 configuration, we observe that the magnetic flux saturates at higher values for lower plasma-$\beta$, consistent with the formation of a MAD state (see Figure \ref{fig:phi_t_LR}). The saturation level for plasma-$\beta=0.1$ transiently reaches the theoretical maximum for MAD \citep{Tchekhovskoy2011} during flux accumulation episodes as seen in Figure \ref{fig:phi_t_LR}, though the time-averaged flux remains approximately half this value due to periodic flux eruption events. This intermittent achievement of the MAD limit suggests that this configuration experiences cycles of magnetic flux buildup and release, characteristic of the MAD state. Nevertheless, for plasma-$\beta=100$ and $1000$ cases, the saturated flux is below the MAD limit, reminiscent of SANE accretion states.
The P2 and PL configurations, however, show a more complex evolution of magnetic flux (see Figure \ref{fig:phi_t_LR}). Interestingly, unlike P1 case, the P2 configuration with plasma-$\beta=100$ exhibits the highest sustained magnetic flux values, frequently reaching and exceeding the MAD saturation threshold, while plasma-$\beta=0.1$ shows lower average flux despite higher variability; plasma-$\beta=1000$ also exhibits average flux below MAD limit. This non-monotonic relationship between initial plasma-$\beta$ and final magnetic flux accumulation suggests that intermediate magnetic field strengths may be optimal for sustained flux accumulation. The PL configuration shows similar complexity as P2, with plasma-$\beta=0.1$ exhibiting intermittent peaks above the MAD threshold but lower time-averaged values compared to plasma-$\beta=100$.

\begin{table}
\centering
\caption{Model properties and accretion states}
\label{tab:models}
\begin{tabular}{lll}
\toprule
Model & Magnetic Flux Properties & Accretion State \\
\midrule
P2B100 & 
\parbox[t]{0.37\textwidth}{Magnetic flux reaches saturation value with time showing a transition to MAD state. Mass accretion shows frequent jumps and variability due to eruption of magnetic flux.} & 
\parbox[t]{0.37\textwidth}{The system acquires enough magnetic flux to form magnetic barriers inducing flux eruption events. This is a MAD state.} \\
\midrule
\parbox[t]{0.25\textwidth}{P1B01, P1B1, PLB100} & 
\parbox[t]{0.37\textwidth}{Magnetic flux reaches saturation value for a brief amount of time and then decays due to flux eruption and thus transitioning to a SANE state. Mass accretion shows variability which are less prominent than the pure MAD state.} & 
\parbox[t]{0.37\textwidth}{The system acquires saturation magnetic flux for some time and transition between SANE and MAD states. This is an intermediate/transition state.} \\
\midrule
\parbox[t]{0.25\textwidth}{P1B10, P1B100, P1B1000, P2B1000, PLB1000} & 
\parbox[t]{0.37\textwidth}{Magnetic flux is always below the saturation value. Mass accretion rate is stable with minimal variability.} & 
\parbox[t]{0.37\textwidth}{The system does not form magnetic barriers and there is steady accretion throughout all time. This is a SANE state.} \\
\midrule
\parbox[t]{0.25\textwidth}{P2B01, P2B1, P2B10, PLB01, PLB1, PLB10} & 
\parbox[t]{0.37\textwidth}{Magnetic flux reaches much above saturation value due to low initial plasma-$\beta$. Mass accretion rate is unstable as the system is disrupted by very high magnetic fields while matter is being thrown out causing a large drop in accretion.} & 
\parbox[t]{0.37\textwidth}{The system forms very strong magnetic barriers and not allowing matter to accrete at all.} \\
\bottomrule
\end{tabular}\\
\vspace{2mm}
{\small For more details about the models see Table~\ref{tab:simulations}.}
\end{table}


\bsp	
\label{lastpage}
\end{document}